\documentclass[nojss,shortnames]{jss}\usepackage[]{graphicx}\usepackage[]{xcolor}
\makeatletter
\def\maxwidth{ %
  \ifdim\Gin@nat@width>\linewidth
    \linewidth
  \else
    \Gin@nat@width
  \fi
}
\makeatother

\usepackage{orcidlink}
\usepackage{thumbpdf}
\usepackage{lmodern}
\usepackage{bold-extra}

\usepackage{amsmath}
\usepackage{amsthm}
\usepackage{amsfonts}
\usepackage{todonotes}
\usepackage{bm}
\usepackage[overload]{empheq}  

\usepackage{float}
\usepackage{tablefootnote}

\theoremstyle{definition}

\newcommand{\continuation}{??}

\author{Leonardo Chiani~\orcidlink{0009-0007-2491-6290}\\Politecnico di Milano 
   \And Emanuele Borgonovo\\Bocconi University
   \AND Elmar Plischke~\orcidlink{0000-0002-2019-9243}\\Helmholtz-Zentrum Dresden-Rossendorf
   \And Massimo Tavoni\\Politecnico di Milano}
\Plainauthor{Leonardo Chiani, Emanuele Borgonovo, Elmar Plischke, Massimo Tavoni}

\title{\pkg{gsaot}: an \proglang{R} package for Optimal Transport-based sensitivity analysis}
\Plaintitle{gsaot: an R package for Optimal Transport-based sensitivity analysis}
\Shorttitle{Optimal Transport-based sensitivity analysis in \proglang{R}}

\Abstract{
  \pkg{gsaot} is an \proglang{R} package for Optimal Transport-based global sensitivity analysis. It provides a simple interface for indices estimation using a variety of state-of-the-art Optimal Transport solvers such as the network simplex and Sinkhorn-Knopp. The package is model-agnostic, allowing analysts to perform the sensitivity analysis as a post-processing step. Moreover, \pkg{gsaot} provides functions for indices and statistics visualization. In this work, we provide an overview of the theoretical grounds, of the implemented algorithms, and show how to use the package in different examples.
}

\Keywords{sensitivity analysis, uncertainty quantification, optimal transport, \proglang{R}}
\Plainkeywords{sensitivity analysis, uncertainty quantification, optimal transport, R}

\Address{
  Leonardo Chiani, Massimo Tavoni\\
  Department of Management Engineering\\
  Politenico di Milano\\
  \emph{and}\\
  CMCC Foundation - Euro-Mediterranean Center on Climate Change, Italy\\
  \emph{and}\\
  RFF-CMCC European Institute on Economics and the Environment, Italy\\
  20133 Milano, Italy\\
  E-mail: \email{leonardo.chiani@polimi.it}, \email{massimo.tavoni@polimi.it}\\
  \medskip
  
  Emanuele Borgonovo\\
  Department of Decision Sciences\\ 
  Bocconi Institute for Data Science and Analytics\\
  Bocconi University\\
  20136 Milano, Italy\\
  E-mail: \email{emanuele.borgonovo@unibocconi.it }\\
  \medskip

  Elmar Plischke\\
  Helmholtz-Zentrum Dresden-Rossendorf\\
  01328 Dresden, Germany\\
  E-mail: \email{e.plischke@hzdr.de}
}
\IfFileExists{upquote.sty}{\usepackage{upquote}}{}
\begin{document}


\section[Introduction]{Introduction} \label{sec:intro}

Sensitivity analysis is pivotal in developing and validating complex models. Through sensitivity analysis, researchers and practitioners understand how the model responds to input variations and which variables drive the model variability \citep{saltelliGlobalSensitivityAnalysis2008}. Its applications have crossed multiple domains in recent years, from finance to climate change mitigation \citep{razaviFutureSensitivityAnalysis2021}. Several libraries implement global sensitivity analysis techniques in different programming languages. \pkg{UQLab} \citep{marelliUQLabFrameworkUncertainty2014} in \proglang{MATLAB}, \pkg{openTURNS} \citep{openTURNS_package} in \proglang{Python}, and \pkg{dakota} \citep{adams2020dakota} in \proglang{C++} include sensitivity analysis methods as part of comprehensive uncertainty quantification pipelines. Looking specifically at global sensitivity analysis, the package \pkg{SALib} \citep{SALib_package_1, SALib_package_2} in \proglang{Python} offers a comprehensive toolkit, including the Sobol' \citep{sobol2001global}, Delta \citep{borgonovo2007new}, and the PAWN methods \citep{pianosi2015simple} among others. Similar comprehensive packages include \pkg{SAFE} \citep{pianosiMatlabToolboxGlobal2015} (developed for \proglang{MATLAB}/\proglang{Octave} and then also translated in \proglang{Python} and \proglang{R}) and \pkg{sensitivity} \citep{sensitivity_package} in \proglang{R}. Other packages, such as \pkg{sensobol} \citep{puySensobolPackageCompute2022} for Sobol' indices and \pkg{multisensi} \citep{multisensi_package} for Sobol' indices for multivariate outputs, provide a clean and simple interface tailored for specific sensitivity indices. 

Most of the previous tools require specific sampling designs or do not address the case of multivariate output. This is a challenging problem for global sensitivity analysis. The growing reliance of scientists on quantitative models and the complexity of research applications has created the need for techniques capable of handling multivariate outputs and correlated inputs. Indeed, multivariate outputs appear, for example, when the outputs are spatially distributed and/or dynamic, as is often the case in physical, climate and socio-economic applications.

At the same time, recent investigations in Statistics and Data Science have renewed attention on measures of statistical association (such as global sensitivity analysis methods), highlighting the importance of properties such as zero-independence and max-functionality \citep{reshef2011detecting,moriFourSimpleAxioms2019a,chatterjeeNewCoefficientCorrelation2021,wieselMeasuringAssociationWasserstein2022}. Zero-independence guarantees that the measure of statistical association between two random variables is equal to zero if and only if the random variables are statistically independent. Max-functionality implies that it reaches its maximum if and only if there is a deterministic functional relationship between the two random variables. Among the few indices that possess these properties, we find Chatterjee's correlation coefficient \citep{chatterjeeNewCoefficientCorrelation2021} and measures of association based on the Wasserstein distance \citep{wieselMeasuringAssociationWasserstein2022}. The Wasserstein distance is grounded in the theory of Optimal Transport \citep{villaniOptimalTransport2009, figalliInvitationOptimalTransport2021}, and the rapid theoretical advances in this area have allowed the formulation of many Optimal Transport-based measures of statistical correlation that handle multivariate outputs and correlated inputs and possess zero-independence and max-functionality \citep{moriEarthMoverCorrelation2020, niesTransportDependencyOptimal2023, borgonovoGlobalSensitivityAnalysis2024}. To our knowledge, the \proglang{R} package \pkg{XICOR} \citep{xicor_package} is the only one implementing one of these methods, the Chatterjee's correlation coefficient. 

Concerning the Optimal Transport problem, many packages provide efficient solvers. In \proglang{Python}, the packages \pkg{POT} \citep{flamary2021pot} and \pkg{OTT} \citep{OTT_package} implement alternative Optimal Transport solvers, such as the network simplex and the Sinkhorn-Knopp algorithms. In \proglang{R}, valid alternatives are the \pkg{T4transport} \citep{T4transport_package} \pkg{approxOT} \citep{approxOT_package}, and \pkg{transport} \citep{transport_package} packages. We selected the latter package because it provides efficient implementations of different solution algorithms for the Kantorovich transport problem and has an easy-to-use interface.

We present an \proglang{R} package that leverages the mathematical framework and the computational advances of Optimal Transport, and the recent results in global sensitivity analysis \citep{borgonovoGlobalSensitivityAnalysis2024}. We implement a given-data estimation procedure that can be employed not only when the simulation model is accessible but also when only a dataset of input-output realizations is available. The key benefits of the resulting approach are:
\begin{itemize}
  \item the capability to handle correlated inputs;
  \item the capability to handle multivariate and complex output data types;
  \item the estimation directly from a dataset; 
  \item a decomposition property of these measures that allocates the change in distribution to the expected impact of a random variable on the first, second, and higher-order statistical moments.
\end{itemize}

The remainder of this paper is organized as follows. In Section \ref{sec:gsaot}, we present the theoretical background of our work. This section is divided into four parts. In Section \ref{subsec:common}, we introduce a common rationale for global sensitivity analysis, providing a framework for our implementation and choice of estimators. We then discuss OT-based sensitivity indices and how they can be defined within such a framework (Sections \ref{subsec:gsaandot} and \ref{subsec:gsaandepsot}). Next, we discuss the computational aspects and their combination with Optimal Transport solvers (Section \ref{subsec:otalgo}). In Section \ref{sec:usage}, we present three applications of \pkg{gsaot}, highlighting the properties of the package. Finally, in Section \ref{sec:summary}, we discuss the main contributions, the limitations, and the scope of our work.

\section{Sensitivity analysis using Optimal Transport} \label{sec:gsaot}

\subsection{Preliminary notions and notation} \label{subsec:common}

We consider two random vectors on $(\Omega, \mathcal B, \Prob)$ with values in $\mathcal X\subseteq \mathbb{R}^d$ and $\mathcal Y\subseteq \mathbb{R}^k$, with $\mathbf X = (X_1, \dots, X_d)$ and $\mathbf Y = (Y_1, \dots, Y_k)$. Let $\mathcal{P}(\mathcal X)$ and $\mathcal{P}(\mathcal Y)$ represent the families of probability laws over $\mathcal X$ and $\mathcal Y$, respectively. We denote with $\Prob_{\mathbf{X}} \in \mathcal{P}(\mathcal X)$ and $\Prob_{\mathbf{Y}} \in \mathcal{P}(\mathcal Y)$ the corresponding probability measures of $\mathbf X$ and $\mathbf Y$. In several applications, analysts need to determine the strength of the statistical association between $\mathbf Y$ and $X_i$. This task is particularly relevant when they are dealing with a mathematical model described by a function $\mathbf{f}\colon \mathcal{X} \subset \mathbb{R}^d \longrightarrow \mathbb{R}^k$, and they need to quantify the relevance of input $X_i$ on the output $\mathbf Y$.

A general way of assessing input importance in global sensitivity analysis is to measure the impact of gathering information about a random input on the output distribution. First, we assess the marginal distribution of the output $\Prob_\mathbf{Y}$. Then, to quantify the effect of receiving information about $X_i$, we fix the value of $X_i$ and compute the conditional distribution of $\mathbf{Y}$ given this information, denoted as $\Prob_{\mathbf{Y} | X_i = x_i}$. Then, we can utilize a separation measurement between $\Prob_\mathbf{Y}$ and $\Prob_{\mathbf{Y} | X_i = x_i}$ to quantify the impact of fixing $X_i$. A separation measurement is a real-valued function $\zeta(\cdot,\cdot)$ defined over the space $\mathcal P (\mathcal Y) \times \mathcal P (\mathcal Y)$, with the property
\begin{equation}
\label{eq:grounded}
\zeta(\Prob, \Prob) = 0 \quad \forall \Prob \in \mathcal P(\mathcal Y)\text{.}
\end{equation}
This requirement ensures that we consistently report no sensitivity if learning $X_i$ has no impact on $\Prob_\mathbf{Y}$, i.e., when $\Prob_\mathbf{Y}=\Prob_{\mathbf{Y}|X_i = x_i}$.
When the quantity of interest $\mathbf{Y}$ is clearly defined in the context, we call the quantity:
\begin{equation}
\label{eq:locsep}
\zeta(x_i)=\zeta(\Prob_\mathbf{Y}, \Prob_{\mathbf{Y}|X_i=x_i}),
\end{equation}
local separation. Formally, $\zeta(x_i)$ is a map from the support of $X_i$ into the real numbers. It quantifies the discrepancy between the output marginal and conditional distribution of $\mathbf{Y}$ given that $X_i$ is fixed at $x_i$.

The common rationale then quantifies the importance of $X_i$ as the expected separation between $\Prob_\mathbf{Y}$, the marginal (unconditional) probability distribution of $\mathbf{Y}$, and the conditional output distribution $\Prob_{\mathbf{Y}|X_i}$:
\begin{equation}
\label{eq:comrat}
    \xi^\zeta (\mathbf{Y}, X_i) = \E[ \zeta(\Prob_{\mathbf{Y}}, \Prob_{\mathbf{Y} | X_i})] = \int
     \zeta(\Prob_{\mathbf{Y}}, \Prob_{\mathbf{Y} | X_i=x_i}) d  \Prob_{X_i}(x_i)\text{.}
\end{equation}
This general framework allows us to obtain a global importance index, as well as regional information on the impact of learning $X_i$ by visualizing the local separation $\zeta(x_i)$ as $X_i$ varies in its support. Many global sensitivity indices and measures of statistical associations are comprised by Equation \eqref{eq:comrat}. Examples are Pearson's correlation coefficient \citep {pearsonGeneralTheorySkew1905}, new Chatterjee's correlation coefficient \citep{chatterjeeNewCoefficientCorrelation2021}, maximum mean discrepancy-based methods \citep{debMeasuringAssociationTopological2020}, and the Wasserstein correlation coefficient of  \cite{wieselMeasuringAssociationWasserstein2022}. This last index is part of the family of OT-based sensitivity indices we discuss in the next section.

\subsection{Optimal Transport-based sensitivity indices} \label{subsec:gsaandot}

The theory of optimal transport (OT) provides the basis for defining a family of metrics over the space of probability distributions with attractive properties. We present the subject concisely, and we refer to \citep{villaniOptimalTransport2009, santambrogioOptimalTransportApplied, peyreComputationalOptimalTransport2020} for an extended treatment of the subject.

Let us consider two probability measures, $\Prob$ and $\Prob'$, defined over the same space $\mathcal{Y}$. The OT problem consists of finding the optimal coupling $\pi^{\ast}$ between $\Prob$ and $\Prob'$ that minimizes a given cost function $c\colon \mathcal{Y} \times \mathcal{Y} \longrightarrow [0, +\infty]$. The couplings or transport plans are joint probability measures on $\mathcal{Y} \times \mathcal{Y}$, such that their marginals are $\Prob$ and $\Prob'$.
In Kantorovich's formulation, the OT problem is stated as:
\begin{equation} 
\label{eq:kantot}
    K(\Prob, \Prob') = \inf_{\pi \in \Pi(\Prob, \Prob')} \int_{\mathcal{Y} \times \mathcal{Y}} c(y, y') \, d\pi(y, y'),
\end{equation}
where $\Pi(\Prob, \Prob')$ is the set of all transport plans. We call $K(\Prob, \Prob')$ the OT cost. In this equation, we can interpret the cost $c(y, y')$ as the effort needed to move one unit of mass from one point $y$ to another point $y'$, and the coupling $\pi(y, y')$ as the amount of mass moved from a point $y$ to $y'$.  

When the cost function is the $p$-th power of a suitable continuous metric, $c(x,y)=d^p(x,y)$, the OT problem defines a metric over the space of distributions called the Wasserstein metric. When $p=2$ and the cost $c$ is the Euclidean distance, \cite{gelbrichFormulaL2Wasserstein1990} proved that the OT solution is given by:
\begin{equation}
\label{eq:wbcomp}
    K(\Prob, \Prob') = \lVert m - m' \rVert^2_2 + \operatorname{Tr} \left( \Sigma + \Sigma' - 2 \left( \Sigma^{\frac{1}{2}} \Sigma' \Sigma^{\frac{1}{2}} \right)^{\frac{1}{2}} \right) + \Gamma(\Prob, \Prob'),
\end{equation}
where $m$, $m'$, and $\Sigma$, $\Sigma'$ are the means and covariance matrices of $\Prob$ and $\Prob'$, respectively, and $\operatorname{Tr}$ is the trace operator. In this expression, the Wasserstein metric is the sum of the cost of matching the first-order moments ($m$ and $m'$), the second-order moments ($\Sigma$ and $\Sigma'$), and all the higher-order moments ($\Gamma(\Prob, \Prob')\geq 0$). In general, $\Gamma(\Prob, \Prob')$ has no analytical solution. However, if $\Prob$ and $\Prob'$ are elliptical distributions with the same characteristic generator, $\Gamma(\Prob, \Prob') = 0$ \citep{gelbrichFormulaL2Wasserstein1990}, and the OT cost is simply the sum of the first two terms in Equation \eqref{eq:wbcomp}. The sum of these two terms is defined in the literature as the Wasserstein-Bures semi-metric \citep{janatiEntropicOptimalTransport2020}, which turns into a metric if we restrict our attention to Gaussian distributions.

Following the framework described in Section \ref{subsec:common}, the cost $K(\Prob, \Prob')$ can be used as a separation measurement (Equation~\eqref{eq:grounded}) to define a measure of statistical association. The OT-based local separation is given by:
\begin{equation}
\label{eq:otlocsep}
    \zeta^{K}(x_i) = K(\Prob_{\mathbf{Y}}, \Prob_{\mathbf{Y} | X_i=x_i}).
\end{equation}
The corresponding measure of statistical association is defined as:
\begin{equation} \label{eq:smot}
  \xi^K (\mathbf{Y}, X_i) = \E[\zeta^K(X_i)] = \E[K(\Prob_{\mathbf{Y}}, \Prob_{\mathbf{Y} | X_i})].
\end{equation}
Furthermore, we can derive a maximum for $\xi^K (\mathbf{Y}, X_i)$:
\begin{equation} \label{eq:upbound}
    \xi^K (\mathbf{Y}, X_i) \leq \mathbb{M}^K [\mathbf{Y}] = \E[c(\mathbf{Y}, \mathbf{Y}')],
\end{equation}
where $\mathbf Y'$ is an independent replica of $\mathbf Y$. Since $\mathbb{M}^{K} [\mathbf{Y}] > 0$, we can further define the OT-based sensitivity index as:
\begin{equation}
\label{eq:otindex}
    \iota^{K} (\mathbf{Y}, X_i) = \frac{\xi^{K} (\mathbf{Y}, X_i)}{\mathbb{M}^K[\mathbf{Y}]}.
\end{equation}
This index has several of the desirable properties studied as axioms for measures of statistical association following \cite{renyiMeasuresDependence1959a} (see also \cite{moriFourSimpleAxioms2019a}) and highlighted in recent works such as  \citet{chatterjeeNewCoefficientCorrelation2021}, \citet{wieselMeasuringAssociationWasserstein2022}, and \citet{borgonovoConvexityMeasuresStatistical2025}:
\begin{enumerate}
  \item Zero-Independence: $\iota^K (\mathbf{Y}, X_i) \geq 0$, and $\iota^K (\mathbf{Y}, X_i) = 0$ if and only if $\mathbf{Y}$ and $X_i$ are independent.
  \item Normalization: $0\leq\iota^K (\mathbf{Y}, X_i) \leq 1$.
  \item Max-functionality: $\iota^K (\mathbf{Y}, X_i) = 1$ if and only if there exists a measurable function $\mathbf{g}$ such that $\mathbf{Y}=\mathbf{g}(X_i)$.
\end{enumerate}
Zero-independence reassures us that a zero value of the measure of association indicates that $\mathbf{Y}$ and $\mathbf{X}$ are statistically independent. Normalization allows us to set the value of the measure of statistical association between zero and unity. Max-functionality specifies that the maximum value (unity) is reached when $\mathbf{Y}$ is a deterministic function of $\mathbf{X}$. In that case, fixing $\mathbf{X}$ at $\mathbf{x}$ makes $\mathbf{Y}$ concentrate on $\mathbf{y}=g(\mathbf{x})$.

In the case of the squared Euclidean cost, using Equation \eqref{eq:wbcomp} as the separation measurement, we can decompose $\iota ^{K}(\mathbf{Y},X_{i})$ into the sum of three terms: 
\begin{equation}
\label{eq:otdecomp}
\iota ^{K}(\mathbf{Y},X_{i})=\iota ^{V}(\mathbf{Y}%
,X_{i})+\iota ^{\Sigma }(\mathbf{Y},X_{i})+\E[\Gamma
(\Prob_{\mathbf{Y}}, \Prob_{\mathbf{Y}|X_{i}})].
\end{equation}%
The first term, called the advective (or variance-based) index, equals 
\begin{equation}
\iota ^{\text{V}}(\mathbf{Y},X_{i})=\dfrac{1}{\mathbb{M}^K[%
\mathbf{Y}]}\sum \limits_{j=1}^{k}\E%
\left[ (\E[Y_{j}]-\E[Y_{j}|X_{i}])^{2}\right] = \dfrac{1}{2}\iota^{LG}(\mathbf{Y},X_{i})\text{.}
\label{eq:var_based}
\end{equation}%
This Equation quantifies the influence of receiving information about $X_{i}$ on the multivariate output means. As expressed in Equation~\eqref{eq:var_based}, $\iota ^{\text{V}}(\mathbf{Y},X_{i})$ equals half the multivariate variance-based sensitivity measure proposed by \cite{lamboniMultivariateSensitivityAnalysis2011} and \cite{gamboaSensitivityAnalysisMultidimensional2014} ($\iota ^{\text{V}}(\mathbf{Y},X_{i})$). The second term, called diffusive (or covariance-based) index, is defined as 
\begin{equation}
\iota ^{\Sigma }(\mathbf{Y},X_{i})=\dfrac{1}{\mathbb{M}^K[Y]}\E\left[ 
\operatorname{Tr}\left( \Sigma _{\mathbf{Y}}+\Sigma _{\mathbf{Y}|X_{i}}-2\left(
\Sigma _{\mathbf{Y}|X_{i}}^{1/2}\Sigma _{\mathbf{Y}}\Sigma _{\mathbf{Y}%
|X_{i}}^{1/2}\right) ^{1/2}\right) \right]   \label{eq:iSigma}
\end{equation}
and represents the impact of knowing $X_{i}$ on the variance-covariance matrix of $\mathbf{Y}$. The third term, $\iota ^{\Gamma }(\mathbf{Y},X_{i})$, is a non-negative remainder that quantifies the effect on $\mathbf{Y}$'s additional higher-order moments. The sum of the first two terms in Equation \eqref{eq:otdecomp} defines the Wasserstein-Bures index 
\begin{equation}\label{eq:iWB}
    \iota ^{WB}(\mathbf{Y},X_{i})=\iota ^{V}(\mathbf{Y},X_{i}) +\iota ^{\Sigma}(\mathbf{Y},X_{i}).
\end{equation}
This index always satisfies  $\iota ^{WB}(\mathbf{Y},X_{i})\leq \iota^K(\mathbf{Y},X_{i})$, and equality holds in the case $\Prob_Y$ and $\Prob_{Y|X_i}$ belong to the family of elliptical distributions with the same characteristic generator \citep{borgonovoGlobalSensitivityAnalysis2024}.

\subsection{Entropic Optimal Transport-based sensitivity indices} \label{subsec:gsaandepsot}

The entropic regularization proposed by \cite{cuturiSinkhornDistancesLightspeed2013} introduces a computationally efficient way to solve OT problems by adding a penalty term to the objective function in Equation \eqref{eq:kantot}. The resulting entropic OT problem is formulated as follows:
\begin{equation} \label{eq:entropicot}
K_{\epsilon}(\Prob, \Prob') = \inf_{\pi \in \Pi(\Prob, \Prob')} \int_{\mathcal{Y} \times \mathcal{Y}} c(x, y) \, d\pi(x, y) + \epsilon \text{KL}(\pi | \Prob \otimes \Prob'),
\end{equation}
where $\epsilon > 0$ is a regularization parameter and $\text{KL}$ denotes the Kullback-Leibler divergence. The regularization makes the problem strongly convex, which permits using the fast Sinkhorn-Knopp algorithm \citep{sinkhornDiagonalEquivalenceMatrices1967}. Moreover, it can be shown that for $\epsilon$ tending to zero, the solution of the entropic formulation tends continuously to the solution of the classical problem in Equation \eqref{eq:kantot} for a broad set of costs \citep{carlier2023convergence}.

We can define the entropic OT-based sensitivity indices as in Equation~\eqref{eq:otindex}:
\begin{equation} \label{eq:eotindex}
   \iota^{K_{\epsilon}} (\mathbf{Y}, X_i) = \frac{\xi^{K_{\epsilon}} (\mathbf{Y}, X_i)}{\mathbb{M}^K[\mathbf{Y}]} = \frac{\E[K_{\epsilon}(\Prob_{\mathbf{Y}}, \Prob_{\mathbf{Y} | X_i})]}{\mathbb{M}^K[\mathbf{Y}]}.
\end{equation}
It can be proven that $\xi^{K_{\epsilon}} \leq \mathbb{M}^{K}[\mathbf{Y}]$, where the maximum value $\mathbb{M}^{K}[\mathbf{Y}]$ is the same as that of the classical OT problem in Equation \eqref{eq:upbound}, and that $\xi^{K_{\epsilon}}(\mathbf{Y}, X_i) \geq K_{\epsilon}(\Prob_{\mathbf{Y}}, \Prob_{\mathbf{Y}}) > 0$ \citep{borgonovoGlobalSensitivityAnalysis2024}. Thus, the entropic OT-based indices have the following properties:
\begin{enumerate}
  \item $\iota^{K_{\epsilon}} (\mathbf{Y}, X_i) \geq \frac{K_{\epsilon}(\Prob_{\mathbf{Y}}, \Prob_{\mathbf{Y}})}{\mathbb{M}^{K}[\mathbf{Y}]} > 0$, and $\iota^{K_{\epsilon}} (\mathbf{Y}, X_i) = \frac{K_{\epsilon}(\Prob_{\mathbf{Y}}, \Prob_{\mathbf{Y}})}{\mathbb{M}^{K}[\mathbf{Y}]}$ if $\mathbf{Y}$ and $X_i$ are independent.
  \item $\iota^{K_{\epsilon}} (\mathbf{Y}, X_i) \leq 1$.
  \item $\iota^{K_{\epsilon}} (\mathbf{Y}, X_i) = 1$ if and only if there exists a measurable function $\mathbf{g}$ such that $\mathbf{Y}=\mathbf{g}(X_i)$.
\end{enumerate}
These indices do not satisfy the zero-independence property. Indeed, cases exist in which $X_i$ and $\mathbf{Y}$ are dependent, yet the lower bound is attained.

These indices can be used in two ways. First, they can be used as well-defined sensitivity indices in their own right with desirable properties, simple interpretation, and fast computation. Second, per the original intuition, we can use them as a computationally fast way to estimate the classical OT-based indices for small values of the regularization parameter $\epsilon$.

\subsection{Computational aspects} \label{subsec:otalgo}

We estimate the OT-based sensitivity indices in Equation \eqref{eq:otindex} using a U-statistic for the denominator $\mathbb M^K[\mathbf Y]$ and given-data estimators \citep{borgonovoCommonRationaleGlobal2016} for the numerator. This approach avoids specific sampling designs and requires only a sample of Monte Carlo simulations. Thus, the computational cost is $N$ model runs. Let $\mathcal{X}_i$ denote the support of the input $X_i$, partitioned into $H$ subsets, $\mathcal{X}_i^h$ for $h \in \{1, \dots, H\}$. A given-data estimator for OT-based sensitivity measures (Equation \eqref{eq:smot}) is defined as:
\begin{equation}
\label{eq:commonestimator}
    \hat{\xi}^K(\mathbf{Y},X_i;N,H) = \frac{1}{H} \sum_{h=1}^H K(\Prob^N_{\mathbf{Y}}, \Prob^N_{\mathbf{Y}|X_i \in \mathcal{X}_i^h}),
\end{equation}
where $\Prob^N$ is the empirical distribution of the output computed from the $N$ realizations, and the condition $X_i = x_i$ is approximated by $X_i \in \mathcal{X}_i^h$. The same idea can be applied to define an estimator for the entropic OT-based indices in Equation \eqref{eq:eotindex}. 

A crucial step in estimating the indices is the calculation of $K(\Prob^N_{\mathbf{Y}}, \Prob^N_{\mathbf{Y}|X_i \in \mathcal{X}_i^h})$, i.e., the numerical solution of the empirical OT problem in Equation \eqref{eq:kantot}. If the output $\mathbf Y$ is one-dimensional ($k=1$), and the cost $c$ is the $L_p^p$ distance, the solution to the empirical OT problem is \citep{peyreComputationalOptimalTransport2020}:
\begin{equation}
  K(\Prob^N_{\mathbf{Y}}, \Prob^N_{\mathbf{Y}|X_i \in \mathcal{X}_i^h}) = \frac{1}{N} \sum_{j=1}^N |\hat{F}^{-1}_{\mathbf{Y}}(\frac{j}{N}) - \hat{F}^{-1}_{\mathbf{Y}|X_i \in \mathcal{X}_i^h}(\frac{j}{N})|^p,
\end{equation}
where $\hat{F}^{-1}_{\mathbf{Y}}$ and $\mathbf Y$ and $\hat{F}^{-1}_{\mathbf{Y}|X_i \in \mathcal{X}_i^h}$ are, respectively, the empirical marginal and conditional quantile functions of $\mathbf{Y}$. The estimation is then fast because it is enough to reorder the marginal and conditional quantiles of $\mathbf Y$ in each partition, calculate the $p$-th power of their differences, and take the average over the partitions.

If the output distributions are elliptical with the same characteristic generator and the cost is the squared Euclidean distance, the solution to the OT problem is:
\begin{equation}
\label{eq:wbest}
\begin{split}
  K(\Prob^N_{\mathbf{Y}}, \Prob^N_{\mathbf{Y}|X_i \in \mathcal{X}_i^h}) = &\sum_{j=1}^{k}
    (\hat{m}_{Y_j} - \hat{m}_{Y_j |X_i \in \mathcal{X}_i^h})^2 + \\
    &\operatorname{Tr} \Bigl( \hat{\Sigma}_{\mathbf{Y}} + \hat{\Sigma}_{\mathbf{Y}|X_i \in \mathcal{X}_i^h} - 2 \bigl( \hat{\Sigma}_{\mathbf{Y}}^{1/2} \hat{\Sigma}_{\mathbf{Y}|X_i \in \mathcal{X}_i^h} \hat{\Sigma}_{\mathbf{Y}}^{1/2} \bigr)^{1/2} \Bigr).
\end{split}
\end{equation}
The estimate in Equation \eqref{eq:wbest} requires only algebraic operations and is therefore computationally fast.
We implemented the ad-hoc functions \code{ot_indices_1d} and \code{ot_indices_wb} for these two cases. Note that the second function can estimate the terms corresponding to first- and second-order moments in Equation \eqref{eq:otdecomp}. 

In the general multivariate output case, the calculation of $K(\Prob^N_{\mathbf{Y}}, \Prob^N_{\mathbf{Y}|X_i \in \mathcal{X}_i^h})$ requires, for each partition $h=1,2,\dots,H$, solving an empirical OT problem of the type:
\begin{equation}
	\begin{array}{c}
		\min_{\mathbf{r}} \frac{1}{N N_h} \sum_{k=1}^{N} \sum_{j:x_{j,i}\in \mathsf{X}_{i}^{h}} r_{k,j} c(y_k,y_j) \\ 
		\text{subject to} \\ 
		\sum_{k=1}^{N}r_{k,j}={N},\sum_{j:x_{j,i}\in 
			\mathsf{X}_{i}^{h}}r_{k,j}={N_{h}},\quad N_{h}=\# \{j:x_{j,i} \in \mathsf{X}_{i}^{h}\} \text{,}%
	\end{array}
	\label{eq:OTGivenData}
\end{equation}%
where $\mathbf{r}$ are the possible empirical
couplings, $\# \{ \cdot \}$ counts the number of elements in a set, $N_{n}$ is the number of
realizations of $X_i$ that fall within the $h$-th partition. The symbols $y_k$ and $y_j$ denote realizations of $\mathbf{Y}$ from $\Prob_{Y}$ and $\Prob_{Y|X\in \mathsf{X}_{i}^{h}}$, respectively.  
\begin{table}[H]
    \centering
    \begin{tabular}{c|c|c|p{13em}}
        \textbf{Name} & \textbf{OT problem} & \textbf{Source package} & \textbf{Reference} \\ \hline
        One-dimensional\tablefootnote{Available only for one-dimensional outputs} & Classic & \pkg{gsaot} & \cite{peyreComputationalOptimalTransport2020} \\
        Wasserstein-Bures\tablefootnote{Available only for squared Euclidean costs} & Classic & \pkg{gsaot} & \cite{gelbrichFormulaL2Wasserstein1990} \\
        Network simplex & Classic & \pkg{transport} & \cite{bonneelDisplacementInterpolationUsing2011a} \\
        Primal-dual & Classic & \pkg{transport} & \cite{luenbergerLinearNonlinearProgramming2021} \\
        Revised simplex & Classic & \pkg{transport} & \cite{luenbergerLinearNonlinearProgramming2021} \\
        Short simplex & Classic & \pkg{transport} & \cite{gottschlichShortlistMethodFast2014} \\
        Sinkhorn & Entropic & \pkg{gsaot} & \cite{cuturiSinkhornDistancesLightspeed2013} \\
        Sinkhorn-stable & Entropic & \pkg{gsaot} & \cite{schmitzerStabilizedSparseScaling2019} \\
    \end{tabular}
    \caption{Algorithms available to estimate $K(\Prob^N_{\mathbf{Y}}, \Prob^N_{\mathbf{Y}|X_i \in \mathcal{X}_i^h})$. \textbf{OT problem} identifies the problem solved (``Classic'' refers to \eqref{eq:kantot}, ``Entropic'' to \eqref{eq:entropicot}), \textbf{Source package}: the \proglang{R} package where the algorithm is implemented, and \textbf{Reference} where it is possible to find an extensive description of the algorithm.}
    \label{tab:otalgo}
\end{table}
Thanks to the rich and recent literature on OT \citep{peyreComputationalOptimalTransport2020}, we implement a wide array of solvers in the function \code{ot_indices}. For the classical OT problem in Equation \eqref{eq:kantot}, we use the \pkg{transport} package as the backend, given the simple interface and the state-of-the-art solvers available. For the empirical entropic OT problem (Equation \eqref{eq:entropicot}), we implement a set of solvers in \proglang{C++} using \pkg{Rcpp} \citep{batesFastElegantNumerical2013} and \pkg{RcppEigen} \citep{eddelbuettelRcppSeamlessIntegration2011}. Table \ref{tab:otalgo} summarizes the algorithms available in \pkg{gsaot}. 

The selection of the partition cardinality $H$ is a delicate step and has been discussed in several works regarding given-data estimation. \cite{strong2013efficient} show that we have a plateau effect for large $N$ and reasonable values of $H$, and the estimate becomes insensitive to the choice of $H$. For instance, at $N=50000$, a selection of $M=50$, or 60 or 70, would not alter the estimates. For sample sizes between 100 and 10000, we suggest a selection of $M$ that ensures at least 100 points per partition set. For small sample sizes, displaying the value of a dummy variable is recommended to check for numerical noise. A dummy variable is an irrelevant auxiliary input to the dataset (we call it a dummy variable as in \cite{noacco2019matlab}) and denote it by $X_{\operatorname{dummy}}$. The intuition is to randomly generate $N$ values of $X_{\operatorname{dummy}}$ independent of the model output $\mathbf Y$. Then, by zero-independence, we have the theoretical value $\iota^K(\mathbf Y,X_{\operatorname{dummy}})=0$. A non-zero estimate of $\iota^K(\mathbf Y,X_{\operatorname{dummy}})$ is the result of numerical noise.  


\section{Examples} \label{sec:usage}

This section illustrates how \pkg{gsaot} works through three use cases. Our first example is a linear Gaussian model for which the ground truth values of the sensitivity indices are known (Subsection \ref{subsec:gaussmodel}). The second use case is the spruce budworm and forest model derived from \cite{puySensobolPackageCompute2022} (Subsection \ref{subsec:spruce+forest}), and the third is a toy climate model (Subsection \ref{subsec:climatemod}). 

The examples start from the Monte Carlo simulation of the models. However, the workflow can be adapted to the other cases by skipping the Monte Carlo simulation part, as \pkg{gsaot} post-processes a given input-output dataset (Subsection \ref{subsec:otalgo}). The generation of this dataset is disentangled from the computation of the sensitivity indices by design. The data-generating process can be a Monte Carlo simulation of a computer code, field experiments, or data retrieval from any other source.

As a preliminary step, we load the packages \pkg{gsaot}, \pkg{ggplot2} \citep{ggplot2_package}, and \pkg{patchwork} \citep{patchwork_package} into the environment using the usual \proglang{R} syntax and set the seed for reproducibility. Indeed, \pkg{gsaot} implements plotting functions based on the \pkg{ggplot2} and \pkg{patchwork}, providing simple interfaces that benefit from the \pkg{ggplot2} grammar of graphics.  
\begin{Schunk}
\begin{Sinput}
R> library("gsaot")
R> library("ggplot2")
R> library("patchwork")
R> set.seed(42)
\end{Sinput}
\end{Schunk}
We present experiments for a linear Gaussian model whose analytical solution is known (Subsection \ref{subsec:gaussmodel}), for the spruce budworm and forest model derived from \cite{puySensobolPackageCompute2022} (Subsection \ref{subsec:spruce+forest}), and for a simple climate model (Subsection \ref{subsec:climatemod}).

\subsection{A linear Gaussian model} \label{subsec:gaussmodel}

The first example is a test case for which the values of OT-based indices are analytically known \citep{borgonovoGlobalSensitivityAnalysis2024}. We use it to explain the properties of some core functions of the \pkg{gsaot} package. 

We consider a bivariate model of the form $\mathbf Y=A\mathbf X^T$, with $\mathbf{Y} = (Y_1, Y_2)$, $\mathbf{X} = (X_1, X_2, X_3)$ and  
\begin{equation*}
A = \begin{bmatrix}
      4 & 2 & 3 \\
      2 & 5 & -1 
    \end{bmatrix}.
\end{equation*}
The inputs follow a multivariate normal distribution, $\mathbf{X} \sim \mathcal{N}(\mathbf{\mu}, \Sigma)$, with $\mu = (1,1,1)$ and 
\begin{equation*}
\Sigma = \begin{bmatrix}
            1 & 0.5 & 0.5 \\
            0.5 & 1 & 0.5 \\
            0.5 & 0.5 & 1
          \end{bmatrix}.
\end{equation*}
\begin{table}[H]
		\centering
		\begin{tabular}{l|c|c|c}
			& $\iota^{V}(\mathbf{Y},X_{i})$ & $\iota^{\Sigma}(\mathbf{Y},X_{i})$ & $\iota^{K}(\mathbf{Y},X_{i})$\\ \hline
			$X_1$ & 0.294 & 0.198 & 0.492 \\
			$X_2$ & 0.318 & 0.189 & 0.507 \\
			$X_3$ & 0.107 & 0.01 & 0.117 \\
		\end{tabular}%
		\caption{Analytical values of $\protect \iota^K(\mathbf{Y},X_{i})$, and associated decomposition into its variance- and covariance-based components for the model in Equation \eqref{eq:otdecomp}.}
        \label{tab:analytical_gauss}
	\end{table}
Table \ref{tab:analytical_gauss} shows the analytical values of the OT-based indices. In this case, the OT-based indices coincide with the corresponding Wasserstein-Bures ones, that is $\protect \iota^K(\mathbf{Y},X_{i})=\protect \iota^{WB}(\mathbf{Y},X_{i})$, as both the marginal and conditional distributions are normal. We use these values to test the estimators presented in Subsection \ref{subsec:otalgo}.

We set the sample size \code{N} to 2000. We then set the number of partitions \code{M} to 20 as a good compromise between the partitioning refinement and the number of elements in each partition. 
\begin{Schunk}
\begin{Sinput}
R> N <- 2000
R> M <- 20
\end{Sinput}
\end{Schunk}
We next define a function that generates the correlated input sample using affine transformations:
\begin{Schunk}
\begin{Sinput}
R> rmvtnorm <- function(N, mu, Sigma) {
+      x1 <- rnorm(N)
+      x2 <- rnorm(N)
+      x3 <- rnorm(N)
+      x <- cbind(x1, x2, x3)
+      x <- x %*% chol(Sigma)
+      x <- sweep(x, 2, mu, "+")
+      colnames(x) <- c("X1", "X2", "X3")
+      return(x)
+  }
\end{Sinput}
\end{Schunk}
We denote with \code{mx} the mean $\mu$ and with \code{Sigmax} the covariance matrix $\Sigma$. We generate the input sample through the following instructions:
\begin{Schunk}
\begin{Sinput}
R> mx <- c(1, 1, 1)
R> Sigmax <- matrix(data = c(1, 0.5, 0.5, 0.5, 1, 0.5,
+      0.5, 0.5, 1), nrow = 3)
R> x <- rmvtnorm(N, mx, Sigmax)
\end{Sinput}
\end{Schunk}
Then, we evaluate the model for each input realisation and store the output in matrix \code{y}:
\begin{Schunk}
\begin{Sinput}
R> A <- matrix(data = c(4, -2, 1, 2, 5, -1), nrow = 2,
+      byrow = TRUE)
R> y <- x %*% t(A)
R> colnames(y) <- c("Y1", "Y2")
\end{Sinput}
\end{Schunk}
We now focus on the sensitivity analysis. The function \code{ot_indices_wb} implements the OT cost estimator in Equation \eqref{eq:wbest} to estimate the Wasserstein-Bures indices (Equation \eqref{eq:iWB}). It takes as arguments the input sample \code{x}, the output sample \code{y}, the number of partitions \code{M}, and a set of arguments for the bootstrap estimation of the indices. 
\begin{Schunk}
\begin{Sinput}
R> indices_wb <- ot_indices_wb(x = x, y = y, M = M)
R> print(indices_wb)
\end{Sinput}
\begin{Soutput}
Method: wass-bures 

Indices:
       X1        X2        X3 
0.4695348 0.4992064 0.1166171 

Advective component:
       X1        X2        X3 
0.2889604 0.3172502 0.1047944 

Diffusive component:
        X1         X2         X3 
0.18057443 0.18195623 0.01182271 
\end{Soutput}
\end{Schunk}
The output of \code{print(indices_wb)} in the previous code section displays the Wasserstein-Bures indices in Equation \eqref{eq:iWB}, evidencing the advective and diffusive components. We identify $X_1$ and $X_2$ as the most important variables. From the values of the advective and diffusive components, we can conclude that their impact is both on the mean and covariance matrix.

We compute the uncertainty on estimating the Wasserstein-Bures indices with the bootstrap confidence intervals, implemented using the \pkg{boot} package. We use 1000 replications and compute the 95\% confidence interval using \code{type="norm"}.
\begin{Schunk}
\begin{Sinput}
R> boot <- TRUE
R> R <- 1000
R> conf <- 0.95
R> type <- "norm"
R> indices_wb <- ot_indices_wb(x = x, y = y, M = M, boot = boot,
+      R = R, conf = conf, type = type)
R> print(indices_wb)
\end{Sinput}
\begin{Soutput}
Method: wass-bures 

Indices:
       X1        X2        X3 
0.4651538 0.4949995 0.1096596 

Advective component:
       X1        X2        X3 
0.2867945 0.3154041 0.1011713 

Diffusive component:
         X1          X2          X3 
0.178359289 0.179595322 0.008488298 

Type of confidence interval: norm 
Number of replicates: 1000 
Confidence level: 0.95 
Bootstrap statistics:
  input  component   original        bias      low.ci    high.ci
1    X1 wass-bures 0.46953482 0.004381033 0.451742695 0.47856489
2    X2 wass-bures 0.49920643 0.004206967 0.481756629 0.50824229
3    X3 wass-bures 0.11661709 0.006957454 0.096203246 0.12311602
4    X1  advective 0.28896039 0.002165893 0.277731348 0.29585766
5    X2  advective 0.31725020 0.001846061 0.307343052 0.32346522
6    X3  advective 0.10479437 0.003623039 0.089842874 0.11249979
7    X1  diffusive 0.18057443 0.002215140 0.172712062 0.18400652
8    X2  diffusive 0.18195623 0.002360906 0.173624053 0.18556659
9    X3  diffusive 0.01182271 0.003334415 0.005728418 0.01124818
\end{Soutput}
\end{Schunk}
The output of \code{print(indices_wb)} with bootstrap contains additional information compared to the previous one. It displays the Wasserstein-Bures indices and their decomposition into advective and diffusive components, and it adds to this results the bootstrap options: type of confidence interval, number of replicates, the confidence level of the corresponding intervals, and the bootstrap statistics. 

We plot the results of the sensitivity analysis using the function \code{plot}. The option \code{wb_all} defines whether or not to plot the advective and diffusive components of the Wasserstein-Bures indices.
\begin{Schunk}
\begin{Sinput}
R> plot(indices_wb, wb_all = TRUE) + scale_fill_brewer(palette = "Dark2")
\end{Sinput}
\begin{figure}[H]
\includegraphics[width=\maxwidth]{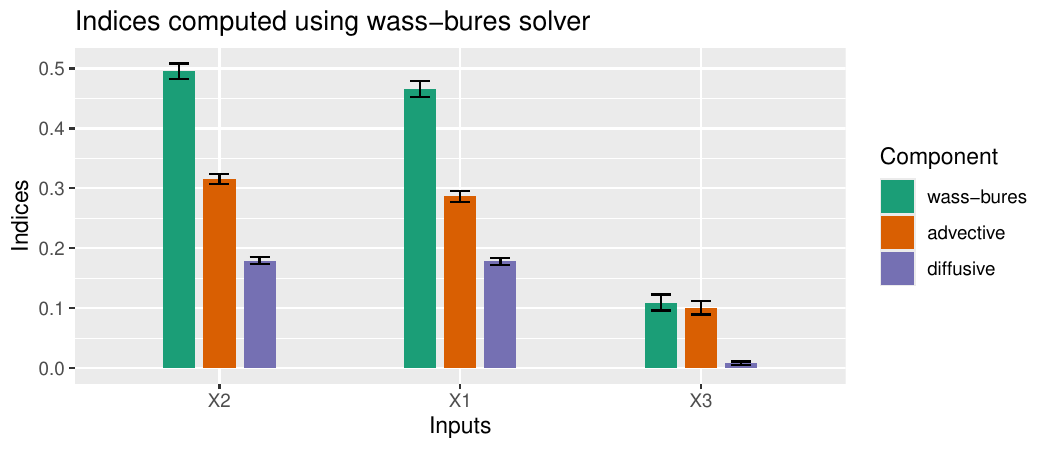} \caption[OT-based indices of the Gaussian model]{OT-based indices of the Gaussian model.}\label{fig:plot_gaussian}
\end{figure}
\end{Schunk}
The bars in Figure \ref{fig:plot_gaussian} represent the values of the corresponding indices, with the error bars indicating the bootstrap confidence intervals.

The Wasserstein-Bures semi-metric is the exact solution to the OT problem for this model. We confirm this result by comparing the indices we obtain from \code{ot_indices_wb} with those obtained from \code{ot_indices} using \code{solver="transport"} and \code{solver="sinkhorn"} with two levels of regularization: large ($\epsilon=0.05$) and small ($\epsilon=0.001$). The argument \code{solver_optns} in \code{ot_indices} allows the user to control the solver's options. For \code{solver = "transport"}, we control the specific numeric solver by changing the \code{method} term. For \code{solver = "sinkhorn"}, we control the regularization with \code{epsilon}, the maximum number of Sinkhorn-Knopp iterations with \code{numIterations}, and the error threshold on the marginals with \code{maxErr}. We compute the different indices with the following instructions:
\begin{Schunk}
\begin{Sinput}
R> indices_wass <- ot_indices(x = x, y = y, M = M, solver = "transport",
+      solver_optns = list(method = "networkflow"))
R> indices_sink_large <- ot_indices(x = x, y = y, M = M,
+      solver = "sinkhorn", solver_optns = list(epsilon = 0.05,
+          numIterations = 1e+06))
R> indices_sink_small <- ot_indices(x = x, y = y, M = M,
+      solver = "sinkhorn", solver_optns = list(epsilon = 0.001,
+          numIterations = 1e+06))
\end{Sinput}
\end{Schunk}
Since \pkg{gsaot} implements various methods for computing the indices, it provides the function \code{plot_comparison} to compare them with a bar plot seamlessly. We use it to compare the four sets of indices computed for the analysed model: \code{indices_wb}, \code{indices_wass}, \code{indices_sink_large}, and \code{indices_sink_small}.
\begin{Schunk}
\begin{Sinput}
R> plot_comparison(list(indices_wb, indices_wass, indices_sink_large,
+      indices_sink_small)) + scale_fill_brewer(palette = "Dark2")
\end{Sinput}
\begin{figure}[H]
\includegraphics[width=\maxwidth]{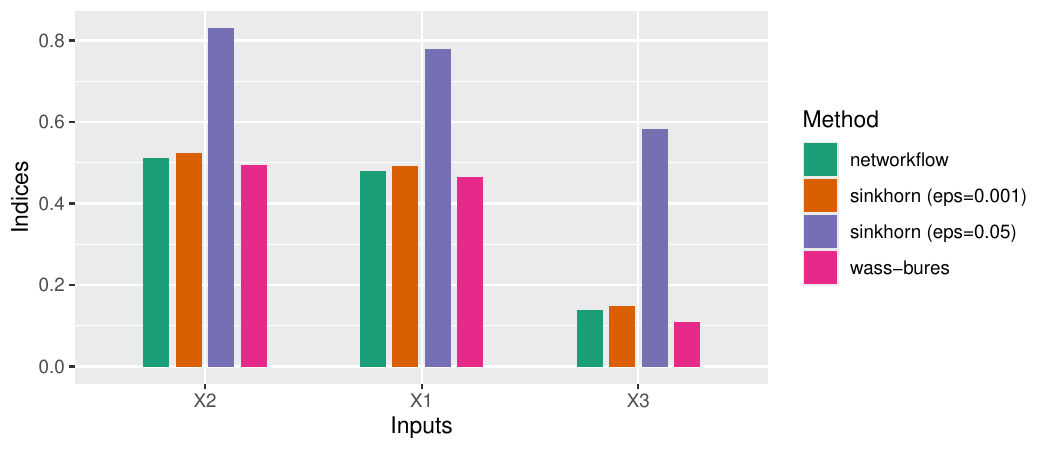} \caption[Comparison of different OT-based indices of the Gaussian model]{Comparison of different OT-based indices of the Gaussian model.}\label{fig:sensitivity_compare_gaussian}
\end{figure}
\end{Schunk}
The results in Figure \ref{fig:sensitivity_compare_gaussian} are consistent with the theory. The Wasserstein-Bures (\code{indices_wb}) and the OT-based indices (\code{indices_wass}) are close to the analytical value in Table \ref{tab:analytical_gauss} for all three variables. Note that these two computational methods rely on entirely different approaches. The Wasserstein-Bures indices are computed using an analytical solution of the OT problem, whereas the second approach uses numerical solvers. The existing gap can be explained by numerical noise and estimation error. Moreover, the entropic OT-based indices with large regularization (\code{indices_sink_large}) correctly identify the importance ranking but are larger than the others by more than 0.25 on average. Conversely, entropic OT-based indices for small regularization parameters (\code{indices_sink_small}) are comparable to the others, confirming the theoretical results on convergence.

We can also compute the importance of the inputs for each singular output ($Y_1$ and $Y_2$) using the function \code{ot_indices_smap}. The function calculates the one-dimensional sensitivity indices for each combination of input and output using the function \code{ot_indices_1d} under the hood. It returns a matrix where each row represents an output and each column represents an input, and the values are the corresponding sensitivity indices. 
\begin{Schunk}
\begin{Sinput}
R> indices_smap <- ot_indices_smap(x, y, M)
R> print(indices_smap)
\end{Sinput}
\begin{Soutput}
          X1         X2        X3
Y1 0.5555770 0.01877416 0.1618665
Y2 0.2954471 0.70408182 0.1040416
\end{Soutput}
\end{Schunk}
The results show that the impact of $X_1$ is higher on $Y_1$, while $X_2$ is very important for $Y_2$ while being negligible for $Y_1$.

\subsection{The spruce budworm and forest model} \label{subsec:spruce+forest}

This example illustrates the use of \pkg{gsaot} in the case of models with time-dependent outputs. The model under scrutiny is the spruce budworm and forest model introduced in \citep{ludwigQualitativeAnalysisInsect2022} and studied in \citep{puySensobolPackageCompute2022}. 

The model consists of the following ordinary differential equations (Equations 20-22 in \cite{ludwigQualitativeAnalysisInsect2022}, see also \cite{puySensobolPackageCompute2022}):
\begin{equation}\label{eq:sprucemodel}
  \left \{
  \begin{array}{l}
  \frac{dB}{dt} = r_B B \left(1 - \frac{B}{KS} \frac{T_E^2 + E^2}{E^2} \right) - \beta \frac{B^2}{(\alpha S)^2 + B^2}\\
  \frac{dS}{dt} = r_S S \left(1 - \frac{SK_E}{EK_S} \right)\\
  \frac{dE}{dt} = r_E E \left(1 - \frac{E}{K_E} \right) - P \frac{B}{S} \frac{E^2}{T_E^2 + E^2}\text{.}
  \end{array}
  \right.
\end{equation}
The model has 10 inputs and 3 time-dependent outputs: $B$, $S$, and $E$. The outputs represent the density of budworms ($B$), the average size ($S$), and the energy reserve of the trees ($E$).
We use the \pkg{deSolve} package \citep{desolve_package} to solve the system of differential equations. 
We consider the time-span $t=1,2,\dots,150$ months with a step of 1 month (\code{times}). The initial conditions are $B_0 = 0.1$, $S_0 = 7$, and $E_0 = 1$ (\code{values0}). We record the output (a collection of 3 time series) for the whole time span. The following code represents the corresponding R instructions.  
\begin{Schunk}
\begin{Sinput}
R> N <- 2000
R> M <- 25
R> times <- seq(0, 150, 1)
R> values0 <- c(B = 0.1, S = 7, E = 1)
\end{Sinput}
\end{Schunk}
We use the \pkg{deSolve} package \cite{desolve_package} to solve the system of differential equations. We define the model using the function obtained from \cite{puySensobolPackageCompute2022}:
\begin{Schunk}
\begin{Sinput}
R> budworm_fun <- function(t, state, parameters) {
+      with(as.list(c(state, parameters)), {
+          dB <- r_b * B * (1 - B/(K * S) * (T_e^2 + E^2)/E^2) -
+              beta * B^2/((alpha^S)^2 + B^2)
+          dS <- r_s * S * (1 - (S * K_e)/(E * K_s))
+          dE <- r_e * E * (1 - E/K_e) - P * (B/S) * E^2/(T_e^2 +
+              E^2)
+          list(c(dB, dS, dE))
+      })
+  }
\end{Sinput}
\end{Schunk}
\begin{table}
    \centering
    \begin{tabular}{c|c|c}
        \textbf{Parameter} & \textbf{Description} & \textbf{Distribution} \\
        \hline
        $r_B$ & Intrinsic budworm growth rate & $\mathcal{U}(1.52, 1.6)$ \\
        $K$ & Maximum budworm density & $\mathcal{U}(100, 355)$ \\
        $\beta$ & Maximum budworm predated & $\mathcal{U}(20000, 43200)$ \\
        $\alpha$ & $\frac{1}{2}$ maximum density for predation & $\mathcal{U}(1, 2)$ \\
        $r_S$ & Intrinsic branch growth rate & $\mathcal{U}(0.095, 0.15)$ \\
        $K_S$ & Maximum branch density & $\mathcal{U}(24000, 25440)$ \\
        $K_E$ & Maximum $E$ level & $\mathcal{U}(1, 1.2)$ \\
        $r_E$ & Intrinsic $E$ growth rate & $\mathcal{U}(0.92, 1)$ \\
        $P$ & Consumption rate of $E$ & $\mathcal{U}(0.0015, 0.00195)$ \\
        $T_E$ & $E$ proportion & $\mathcal{U}(0.7, 0.9)$ \\
    \end{tabular}
    \caption{Input distributions in the spruce budworm and forest model \citep{puySensobolPackageCompute2022}}
    \label{tab:spruceforest_inputs}
\end{table}
For a thorough comparison, we select the same input distributions as in \cite{puySensobolPackageCompute2022} for our numerical experiments (Table \ref{tab:spruceforest_inputs}). Similarly to Subsection \ref{subsec:gaussmodel}, we create the input sample using the custom function:
\begin{Schunk}
\begin{Sinput}
R> budworm_input <- function(N) {
+      params <- data.frame(r_b = runif(N, 1.52, 1.6),
+          K = runif(N, 100, 355), beta = runif(N, 20000,
+              43200), alpha = runif(N, 1, 2), r_s = runif(N,
+              0.095, 0.15), K_s = runif(N, 24000, 25440),
+          K_e = runif(N, 1, 1.2), r_e = runif(N, 0.92,
+              1), P = runif(N, 0.0015, 0.00195), T_e = runif(N,
+              0.7, 0.9))
+      return(params)
+  }
R> x <- budworm_input(N)
\end{Sinput}
\end{Schunk}
We implement parallelization using \pkg{future} \citep{future_package} and \pkg{future.apply} \citep{furrr_package}. We load the packages and we set the backend to \code{multisession}, with the number of workers (processes in parallel) equal to 50\% of available cores.
\begin{Schunk}
\begin{Sinput}
R> library("future")
R> library("future.apply")
R> plan(multisession(workers = availableCores() * 0.5))
\end{Sinput}
\end{Schunk}
Next, we next solve the ODE system for each row of \code{x}, and we gather the outputs in a \code{data.frame} called \code{y}:
\begin{Schunk}
\begin{Sinput}
R> library(deSolve)
R> y <- future_apply(X = x, MARGIN = 1, ode, y = values0,
+      times = times, func = budworm_fun)
R> y <- as.data.frame(y)
\end{Sinput}
\end{Schunk}
We label the output rows:
\begin{Schunk}
\begin{Sinput}
R> y[, "variable"] <- rep(c("time", "B", "S", "E"), each = length(times))
\end{Sinput}
\end{Schunk}
We perform the analysis separately for the three time series. To do so, we store each output in a matrix with rows corresponding to different input combinations and columns corresponding to the time steps:
\begin{Schunk}
\begin{Sinput}
R> B <- as.matrix(t(y[y[, "variable"] == "B", -ncol(y)]))
R> colnames(B) <- times
R> S <- as.matrix(t(y[y[, "variable"] == "S", -ncol(y)]))
R> colnames(S) <- times
R> E <- as.matrix(t(y[y[, "variable"] == "E", -ncol(y)]))
R> colnames(E) <- times
\end{Sinput}
\end{Schunk}
Then, we perform the sensitivity analysis on each output curve separately. We choose as solver \code{"transport"} and as numerical method \code{"networkflow"} as in Subsection \ref{subsec:gaussmodel}. We compute the indices with the function \code{ot_indices}:
\begin{Schunk}
\begin{Sinput}
R> solver <- "transport"
R> solver_optns <- list(method = "networkflow")
R> indices_B <- ot_indices(x = x, y = B, M = M, solver = solver,
+      solver_optns = solver_optns)
R> indices_S <- ot_indices(x = x, y = S, M = M, solver = solver,
+      solver_optns = solver_optns)
R> indices_E <- ot_indices(x = x, y = E, M = M, solver = solver,
+      solver_optns = solver_optns)
\end{Sinput}
\end{Schunk}
We focus on output $B$ to highlight the problem of numerical noise in the OT-based indices estimation. The OT-based indices are reported below:
\begin{Schunk}
\begin{Sinput}
R> print(indices_B)
\end{Sinput}
\begin{Soutput}
Method: transport 

Indices:
       r_b          K       beta      alpha        r_s        K_s 
0.02905249 0.58258164 0.02830713 0.02642674 0.21522521 0.02893317 
       K_e        r_e          P        T_e 
0.04078548 0.02592461 0.03021724 0.03136537 
\end{Soutput}
\end{Schunk}
The values of the OT-based indices reveal $K$ and $r_S$ as important inputs, while the remaining inputs have OT-based indices close to zero. The problem of whether these inputs are not influential is transversal to numerical applications and boils down to understanding whether the non-null estimates are due to numerical noise. The estimate $\widehat{\iota}^K(\mathbf Y,X_{\operatorname{dummy}})$ introduced in Subsection \ref{subsec:otalgo} can then be used as a threshold to define a potentially irrelevant input. In fact, if we record $\widehat{\iota}^K(\mathbf Y,X_{i}) \approx \widehat{\iota}^K(\mathbf Y,X_{\operatorname{dummy}})$, then $X_{i}$ can be considered close to an irrelevant variable. We call $\widehat{\iota}^K(\mathbf Y,X_{\operatorname{dummy}})$ irrelevance threshold. In \code{gsaot}, \code{irrelevance_threshold} performs this function. It allows the user to specify the solver and alternative distributions of the dummy. In our example, we select a uniform distribution and we compute the dummy variables for the three outputs considered
\begin{Schunk}
\begin{Sinput}
R> dummy_optns <- list(distr = "runif")
R> dummy_B <- irrelevance_threshold(B, M, dummy_optns = dummy_optns,
+      solver = solver, solver_optns = solver_optns)
R> dummy_S <- irrelevance_threshold(S, M, dummy_optns = dummy_optns,
+      solver = solver, solver_optns = solver_optns)
R> dummy_E <- irrelevance_threshold(E, M, dummy_optns = dummy_optns,
+      solver = solver, solver_optns = solver_optns)
\end{Sinput}
\end{Schunk}
The function \code{plot.gsaot_indices} allows the user to provide a dummy variable to be added to the indices barplot using the argument \code{threshold}. We then plot the indices and the irrelevance threshold:
\begin{Schunk}
\begin{Sinput}
R> ((plot(indices_B, threshold = dummy_B) +
+     labs(title = "Output B")) / 
+     (plot(indices_S, threshold = dummy_S) +
+     labs(title = "Output S")) / 
+     (plot(indices_E, threshold = dummy_E) +
+     labs(title = "Output E"))) +
+    plot_layout(axes = "collect") &
+    scale_fill_brewer(palette = "Dark2") &
+    scale_y_continuous(limits = c(0, 0.85))
\end{Sinput}
\begin{figure}[H]
\includegraphics[width=\maxwidth]{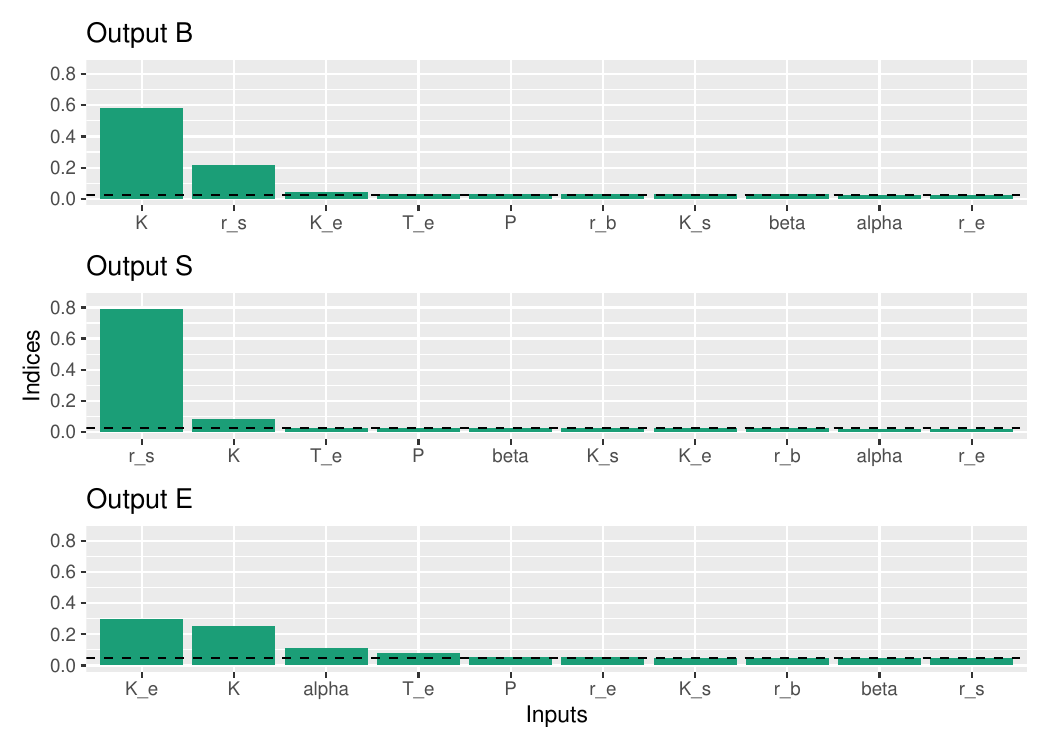} \caption[OT-based indices of the spruce buworm and forest model]{OT-based indices of the spruce buworm and forest model.}\label{fig:plot_sprucebudworm}
\end{figure}
\end{Schunk}
Figure \ref{fig:plot_sprucebudworm} shows the indices and the irrelevance threshold. Two inputs are predominantly relevant for output $B$: $K$ and $r_S$. A third input, $T_e$, has an OT-based index slightly higher than the irrelevance threshold. All the other indices have values smaller than the irrelevance threshold. Similarly, the output $S$ depends only on $r_S$ and $K$, with a possibly small contribution from $T_e$. The output $E$ depends on $K_E$, $K$, $\alpha$, and $T_e$. Figure \ref{fig:plot_sprucebudworm} shows results similar to the ones in Figure 16, Section 3 of \cite{puySensobolPackageCompute2022} but allows us to define a clear and unique importance ranking with fewer model runs.

To gain further insights into how the inputs affect the output, \code{gsaot} allows the user to plot the OT-based local separations (Equation~\eqref{eq:otlocsep}) of the inputs using the function \code{plot_separations}. \code{plot_separations} plots the scaled local separations for the partitions $\mathcal{X}_i^h$ for $h \in \{1, \dots, H\}$, defined as $\zeta^K(x_i) \mathbb{M}^K[\mathbf{Y}]^{-1}$. The optional argument \code{ranking} allows the user to select the inputs to plot according to the ranking. We plot the local separations of the three most important inputs for output $B$. 
\begin{Schunk}
\begin{Sinput}
R> plot_separations(indices_B, ranking = 3) & scale_y_continuous(limits = c(0,
+      1.6))
\end{Sinput}
\begin{figure}[H]
\includegraphics[width=\maxwidth]{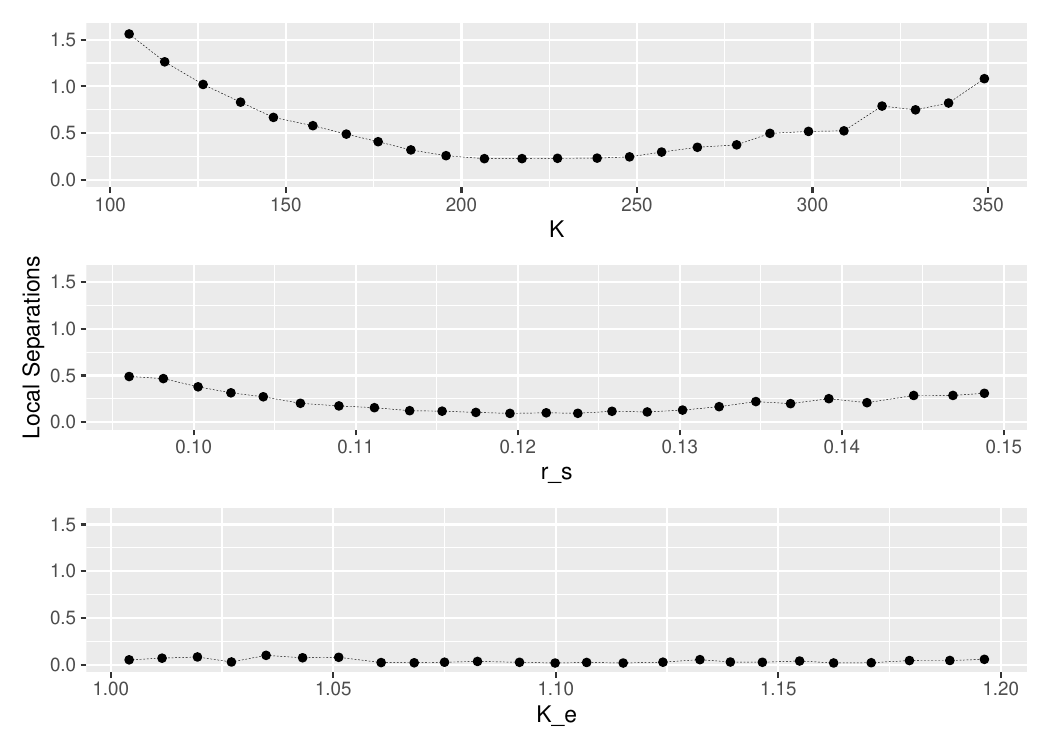} \caption[Local separations of the three most relevant inputs in the spruce budworm and forest model]{Local separations of the three most relevant inputs in the spruce budworm and forest model.}\label{fig:plot_is_sprucebudworm}
\end{figure}
\end{Schunk}
Figure \ref{fig:plot_is_sprucebudworm} reports the input values on the x-axis and the corresponding local separation, scaled by the upper bound (Equation \eqref{eq:upbound}), on the y-axis. The figure reveals that $B$ is primarily influenced by extreme values of $K$ and $r_S$, especially at low values. In contrast, the central values of both parameters have a minimal effect. Additionally, the local separations for $T_e$ confirm the insignificance of this input, showing no regions where it has a substantial impact.

\subsection{A climate model} \label{subsec:climatemod}

This last test case focuses on the implementation of custom ground costs. In some circumstances, analysts may want to change the ground cost to better reflect the structure of the output space. To do so, the package allows users to customize the ground cost $c$.

To illustrate the procedure, we consider the climate module implemented in the 2016 release of DICE \citep{nordhausRevisitingSocialCost2017}. These equations represent the dynamics of CO\textsubscript{2} concentration among different reservoirs and the resulting changes in global temperature. The DICE model uses these equations to simulate and analyze the impact of policy decisions on climate change and the economy. The following discrete-time system defines the model:
\begin{align*}
M_{AT}(t+1) &= \varphi_{11} M_{AT}(t) + \varphi_{21} M_{UO}(t) + 5 E(t) \\
M_{UO}(t+1) &= \varphi_{12} M_{AT}(t) + \varphi_{22} M_{UO}(t) + \varphi_{32} M_{LO}(t) \\
M_{LO}(t+1) &= \varphi_{23} M_{UO}(t) + \varphi_{33} M_{LO}(t) \\
T_{AT}(t+1) &= T_{AT}(t) + c_1 \left( F(t+1) - \frac{\lambda}{S} T_{AT}(t) - c_3 (T_{AT}(t) - T_{OC}(t)) \right) \\
T_{OC}(t+1) &= T_{OC}(t) + c_4 \left( T_{AT}(t) - T_{OC}(t) \right) \\
F(t) &= \lambda \ln \left( \frac{M_{AT}(t)}{M_{AT, pre}} \right) + F_{EX}(t) \\
F_{EX}(t) &= F_{EX0} + \frac{1}{17} (F_{EX1} - F_{EX0}) (t - 1)
\end{align*}
The model has 16 inputs and calculates six outputs. Among the inputs, five are functionally dependent, while one (the annual emissions $E(t)$) is endogenous in the DICE model. In this experiment, we run the DICE2016 model (publicly available at \url{https://yale.app.box.com/s/whlqcr7gtzdm4nxnrfhvap2hlzebuvvm/folder/178027653846}) in its base form and extract the CO\textsubscript{2} emissions results from 2015 until 2100. 
\begin{table}[h!]
\centering
\begin{tabular}{c|c|c}
\textbf{Parameter} & \textbf{Description} & \textbf{Distribution Type} \\ \hline
$\varphi_{11}$ & Carbon cycle transition matrix & $\mathcal{U}(0.704, 1.056)$ \\ 
$\varphi_{23}$ & Carbon cycle transition matrix & $\mathcal{U}(0.0056, 0.0084)$ \\ 
$c_{1}$ & Climate equation coefficient for upper level & $\mathcal{U}(0.0804, 0.1206)$ \\ 
$c_{3}$ & Transfer coefficient upper to lower stratum & $\mathcal{U}(0.0704, 0.1056)$ \\
$c_{4}$ & Transfer coefficient for lower level & $\mathcal{U}(0.0.02, 0.03)$ \\ 
$\lambda$ & Estimated forcings of equilibrium CO2 doubling & $\mathcal{U}(2.94504, 4.41756)$ \\ 
$S$ & Climate Sensitivity & $\mathcal{U}(2.48, 3.72)$ \\ 
$F_{EX0}$ & 2015 forcings of non-CO2 GHG (Wm-2) & $\mathcal{U}(0.4, 0.6)$ \\ 
$F_{EX1}$ & 2100 forcings of non-CO2 GHG (Wm-2) & $\mathcal{U}(0.8, 1.2)$ \\ 
\end{tabular}
\caption{Input distributions of the climate model}
\label{tab:param_distributions}
\end{table}

Table \ref{tab:param_distributions} displays the distribution of the model inputs of the model, derived from \cite{andersonUncertaintyClimateChange2014}. Concerning the outputs, we focus on atmospheric temperature anomaly $T_{AT}$ for its importance in policy-making. We inizialize the simulation using the default DICE2016 values.
In the code implementation, we build a function that receives as input the emissions curve $\{E(t)\}_t$, the parameters of the model and the initial states and returns the atmospheric temperature $T_{AT}$. The \proglang{R} code for the model is:
\begin{Schunk}
\begin{Sinput}
R> climate_model <- function(emissions, params, init) {
+      with(as.list(c(params, init)), {
+          phi12 <- 1 - phi11
+          phi21 <- phi12 * 588/360
+          phi22 = 1 - phi21 - phi23
+          phi32 = phi23 * 360/1720
+          phi33 = 1 - phi32
+          M_AT <- M_AT_init
+          M_UO <- M_UO_init
+          M_LO <- M_LO_init
+          T_AT <- T_AT_init
+          T_OC <- T_OC_init
+          T_AT_values <- numeric(length(emissions))
+          T_AT_values[1] <- T_AT
+          for (t in 1:(length(emissions) - 1)) {
+              E <- emissions[t]
+              M_AT_next <- phi11 * M_AT + phi21 * M_UO +
+                  5 * E
+              M_UO_next <- phi12 * M_AT + phi22 * M_UO +
+                  phi32 * M_LO
+              M_LO_next <- phi23 * M_UO + phi33 * M_LO
+              F_EX <- F_EX0 + (F_EX1 - F_EX0) * (t -
+                  1)/17
+              F_next <- lambda * log(M_AT/588)/log(2) +
+                  F_EX
+              T_AT_next <- T_AT + c1 * (F_next - lambda *
+                  T_AT/S - c3 * (T_AT - T_OC))
+              T_OC_next <- T_OC + c4 * (T_AT - T_OC)
+              T_AT_values[t + 1] <- T_AT_next
+              M_AT <- M_AT_next
+              M_UO <- M_UO_next
+              M_LO <- M_LO_next
+              T_AT <- T_AT_next
+              T_OC <- T_OC_next
+          }
+          return(T_AT_values)
+      })
+  }
\end{Sinput}
\end{Schunk}
In the following code, we instantiate the emissions and initial values.
\begin{Schunk}
\begin{Sinput}
R> emissions <- c(35.74, 33.22, 35.26, 36.96, 38.28, 39.19,
+      39.67, 39.7, 39.29, 38.43, 37.12, 35.38, 33.22,
+      30.65, 27.71, 24.4, 20.76, 16.82)/3.666
R> init <- data.frame(M_AT_init = 851, M_UO_init = 460,
+      M_LO_init = 1740, T_AT_init = 0.85, T_OC_init = 0.0068)
\end{Sinput}
\end{Schunk}
For this example, we run 2000 model simulations, and we fix the number of partitions to 15. 
\begin{Schunk}
\begin{Sinput}
R> N <- 2000
R> M <- 15
\end{Sinput}
\end{Schunk}
As in the previous examples, we define a function that prepares the input sample: 
\begin{Schunk}
\begin{Sinput}
R> sample_parameters <- function(N) {
+    params <- data.frame(
+      phi11 = runif(N, 0.704, 1.056),
+      phi23 = runif(N, 0.0056, 0.0084),
+      c1 = runif(N, 0.0804, 0.1206),
+      c3 = runif(N, 0.0704, 0.1056),
+      c4 = runif(N, 0.02, 0.03),
+      lambda = runif(N, 2.94504, 4.41756),
+      S = runif(N, 2.48, 3.72),
+      F_EX0 = runif(N, 0.4, 0.6),
+      F_EX1 = runif(N, 0.8, 1.2)
+    )
+    return(params)
+  }
R> x <- sample_parameters(N)
\end{Sinput}
\end{Schunk}
We use the \code{apply} function to run the model in an efficient way over the input \code{data.frame} \code{x}.
\begin{Schunk}
\begin{Sinput}
R> y <- apply(X = x, MARGIN = 1, climate_model, emissions = emissions,
+      init = init)
R> y <- t(y)
\end{Sinput}
\end{Schunk}
\begin{Schunk}
\begin{figure}[H]

{\centering \includegraphics[width=\maxwidth]{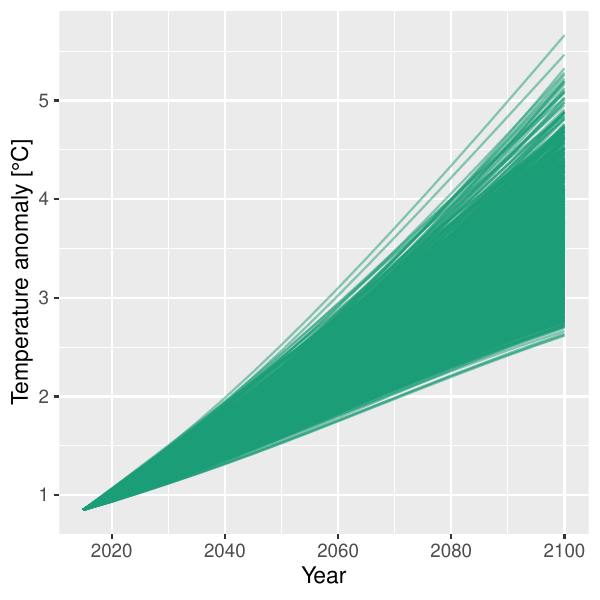} 

}

\caption[Temperature anomaly trajectories of the analyzed climate model]{Temperature anomaly trajectories of the analyzed climate model}\label{fig:plot_distribution_dice}
\end{figure}
\end{Schunk}
Figure \ref{fig:plot_distribution_dice} displays a sample of output trajectories of the model. To calculate the global sensitivity indices, we use the fourth power of Minkowski distance of order 4 ($L_3^3$) as ground cost, implemented through the \code{stats::dist} function. We wrap it into the function \code{custom_metric} that receives as input a matrix \code{y} where rows are data points and returns as output a cost matrix where the element in position $(i,j)$ is the cost related to rows $i$ and $j$ in \code{y}:
\begin{Schunk}
\begin{Sinput}
R> custom_metric <- function(y) {
+      cost_matrix <- as.matrix(stats::dist(y, method = "minkowski",
+          p = 3))^3
+      return(cost_matrix)
+  }
\end{Sinput}
\end{Schunk}
We then compute the entropic OT-based indices. We use the obtained indices as a proxy of the slower simplex solver, as discussed in Subsection \ref{subsec:gsaandepsot}. We use \code{solver_optns} to control the regularisation (\code{epsilon}), increase the maximum number of iterations (\code{numIterations}), and set the error tolerance (\code{maxErr}). We also compute the 95\% confidence intervals for the indices using 100 bootstrap replicates. We also parallelize the bootstrap computations using 75\% of the available cores using the arguments \code{parallel} and \code{ncpus}.
\begin{Schunk}
\begin{Sinput}
R> solver <- "sinkhorn"
R> solver_optns <- list(epsilon = 0.001, numIterations = 1e+05,
+      maxErr = 0.001)
R> boot <- TRUE
R> R <- 100
R> parallel <- "multicore"
R> ncpus <- 0.75 * availableCores()
R> conf <- 0.95
R> type <- "basic"
\end{Sinput}
\end{Schunk}
We then use the function \code{ot_indices} as before, but we specify the custom function using the \code{cost} argument.
\begin{Schunk}
\begin{Sinput}
R> indices_sink <- ot_indices(x, y, M, cost = custom_metric,
+      solver = solver, solver_optns = solver_optns, boot = boot,
+      R = R, parallel = parallel, ncpus = ncpus, conf = conf,
+      type = type)
R> print(indices_sink)
\end{Sinput}
\begin{Soutput}
Method: sinkhorn 

Indices:
     phi11      phi23         c1         c3         c4     lambda 
0.21173030 0.02372565 0.04033141 0.02220658 0.02290184 0.02833276 
         S      F_EX0      F_EX1 
0.07917945 0.02202884 0.02445545 

Type of confidence interval: basic 
Number of replicates: 100 
Confidence level: 0.95 
Bootstrap statistics:
   input   original        bias     low.ci    high.ci
1  phi11 0.21809242 0.006362116 0.19061732 0.23182383
2  phi23 0.02716026 0.003434608 0.01995295 0.02803990
3     c1 0.04518088 0.004849468 0.03420690 0.04583295
4     c3 0.02554771 0.003341124 0.01882086 0.02468860
5     c4 0.02661961 0.003717774 0.01893716 0.02593977
6 lambda 0.03225555 0.003922790 0.02315187 0.03358493
7      S 0.08475171 0.005572263 0.06515390 0.09085698
8  F_EX0 0.02556509 0.003536249 0.01769989 0.02514948
9  F_EX1 0.02808882 0.003633370 0.02091851 0.02787862
\end{Soutput}
\end{Schunk}
From the results displayed by \code{print(indices_sink)}, we identify four relevant inputs: $\varphi_{11}$, $S$, $c_1$, and $\lambda$. On the other hand, the OT-based indices of variables $\varphi_{23}$, $c_3$, $c_4$, $F_{EX0}$, and $F_{EX1}$ are small. We compute the irrelevance thresholds to confirm that these inputs are unimportant using the dummy approach described in Section \ref{subsec:spruce+forest}. 
\begin{Schunk}
\begin{Sinput}
R> dummy_dice <- irrelevance_threshold(y, M, cost = custom_metric,
+      solver_optns = solver_optns)
R> print(dummy_dice)
\end{Sinput}
\begin{Soutput}
Method: sinkhorn 

Indices:
     rnorm 
0.02646428 
\end{Soutput}
\end{Schunk}
We then plot the indices with the irrelevance threshold as before: 
\begin{Schunk}
\begin{Sinput}
R> plot(indices_sink, threshold = dummy_dice) +
+    scale_fill_brewer(palette = "Dark2")
\end{Sinput}
\begin{figure}[H]
\includegraphics[width=\maxwidth]{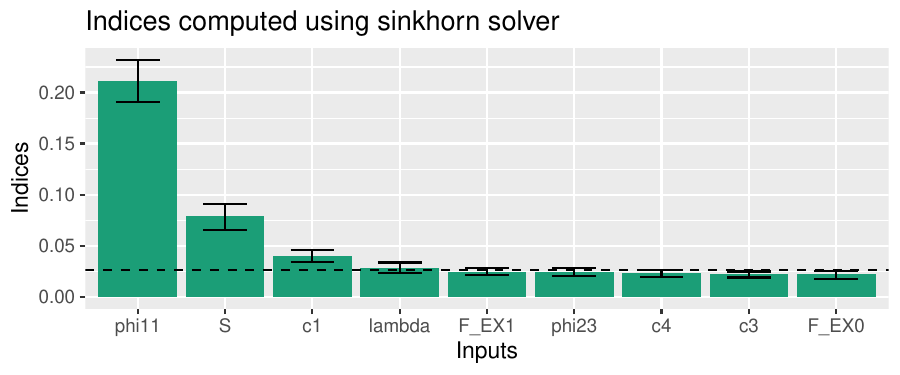} \caption[OT-based indices of the climate model]{OT-based indices of the climate model.}\label{fig:plot_dice}
\end{figure}
\end{Schunk}
We explore further the low-importance inputs by looking at their local separations as in Subsection \ref{subsec:spruce+forest}. This analysis can be done using the \code{ranking} argument of \code{plot_separations}. With \code{ranking=-3}, the function returns a \code{patchwork} object representing the separation measures of the three least important inputs. Moreover, using \pkg{patchwork} functions, we can further manipulate the figure to highlight the zero.  
\begin{Schunk}
\begin{Sinput}
R> plot_separations(indices_sink, ranking = -3) & geom_hline(yintercept = 0,
+      linetype = 2, alpha = 0.6)
\end{Sinput}
\begin{figure}
\includegraphics[width=\maxwidth]{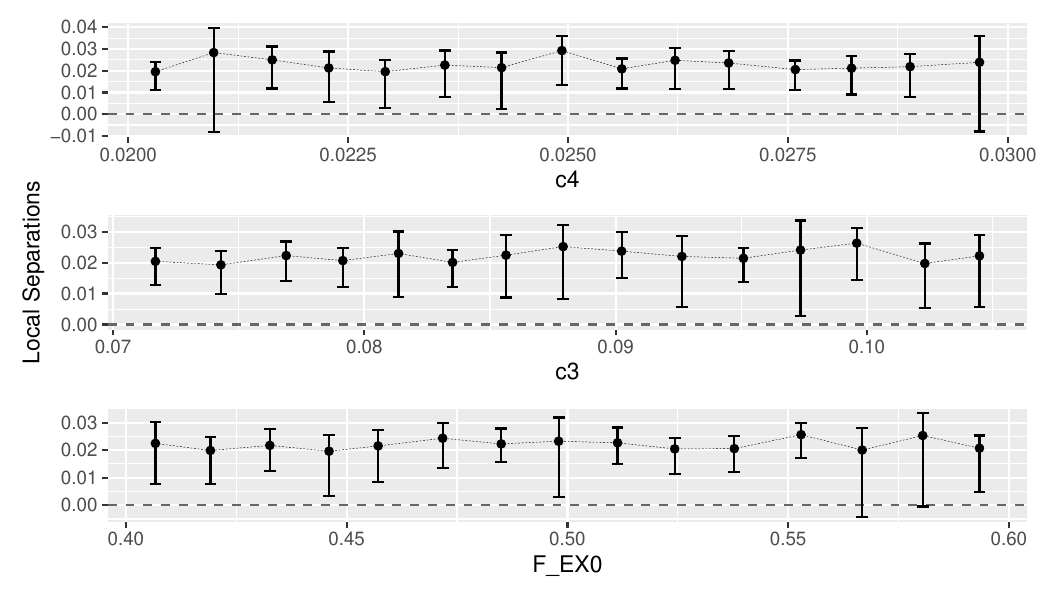} \caption[Local separations of the last three inputs per importance in the climate model]{Local separations of the last three inputs per importance in the climate model.}\label{fig:plot_is_dice}
\end{figure}
\end{Schunk}
The plot shows that all the inputs considered ($phi_{23}$, $F_{EX1}$, and $c_4$) have a similarly small effect in the whole domain. Using this result, we can safely decide to fix these inputs for future model analyses.


\section{Discussion and conclusions} \label{sec:summary}

Sensitivity analysis is an invaluable tool in a world where complex mathematical models support critical decision-making processes. The \pkg{gsaot} package implements a robust approach based on the Optimal Transport theory, offering a powerful and flexible framework for assessing input importance in black-box models. Because the analysis is entirely based on post-processing a given dataset, users can run their models in any programming language.

The sensitivity indices estimated by \pkg{gsaot} are measures of statistical association that possess notable properties such as zero-independence and max-functionality. Moreover, these indices connect measures of associations based on variance contribution (i.e., Sobol' indices) with the family of moment-independent indices and are well-suited to address multivariate outputs. They are also well defined in the presence of correlated inputs.

Thus, \pkg{gsaot} is, to our knowledge, the first package that can handle both multivariate outputs and correlated inputs with a given-data approach. This flexibility differentiates \pkg{gsaot} from well-established tools such as \pkg{sensobol} and \pkg{sensitivity}. Furthermore, \pkg{gsaot} offers significant computational efficiency and flexibility advantages. It achieves high performance by leveraging \proglang{C++} via the \pkg{transport} package and the \pkg{Rcpp} interface. It also accommodates any Monte Carlo sampling design and supports diverse input data types, making it highly versatile. 

\pkg{gsaot} is available on the Comprehensive \proglang{R} Archive Network (CRAN) at \url{https://cran.r-project.org/package=gsaot} or on GitHub at \url{https://github.com/pietrocipolla/gsaot}. 


\section*{Acknowledgments}

L.C. and M.T. acknowledge funding from the European Union’s European Research Council (ERC), Grant project No 101044703 (EUNICE). E.P. acknowledges funding by the German Federal Ministry for the Environment, Nature Conservation, Nuclear Safety and Consumer Protection (BMUV) under grant \#02E21112B.


\bibliography{gsaot_jss_compliant,packages}

\begin{thebibliography}{59}
\newcommand{\enquote}[1]{``#1''}
\providecommand{\natexlab}[1]{#1}
\providecommand{\url}[1]{\texttt{#1}}
\providecommand{\urlprefix}{URL }
\expandafter\ifx\csname urlstyle\endcsname\relax
  \providecommand{\doi}[1]{doi:\discretionary{}{}{}#1}\else
  \providecommand{\doi}{doi:\discretionary{}{}{}\begingroup \urlstyle{rm}\Url}\fi
\providecommand{\eprint}[2][]{\url{#2}}

\bibitem[{Adams \emph{et~al.}(2020)Adams, Bohnhoff, Dalbey, Ebeida, Eddy, Eldred, Hooper, Hough, Hu, Jakeman \emph{et~al.}}]{adams2020dakota}
Adams BM, Bohnhoff WJ, Dalbey KR, Ebeida MS, Eddy JP, Eldred MS, Hooper RW, Hough PD, Hu KT, Jakeman JD, \emph{et~al.} (2020).
\newblock \enquote{\pkg{Dakota}, a multilevel parallel object-oriented framework for design optimization, parameter estimation, uncertainty quantification, and sensitivity analysis: version 6.13 user's manual.}
\newblock \emph{Technical report}, Sandia National Lab.(SNL-NM), Albuquerque, NM (United States).

\bibitem[{Anderson \emph{et~al.}(2014)Anderson, Borgonovo, Galeotti, and Roson}]{andersonUncertaintyClimateChange2014}
Anderson B, Borgonovo E, Galeotti M, Roson R (2014).
\newblock \enquote{Uncertainty in {{Climate Change Modeling}}: {{Can Global Sensitivity Analysis Be}} of {{Help}}?}
\newblock \emph{Risk Analysis}, \textbf{34}(2), 271--293.
\newblock ISSN 1539-6924.
\newblock \doi{10.1111/risa.12117}.

\bibitem[{Bates and Eddelbuettel(2013)}]{batesFastElegantNumerical2013}
Bates D, Eddelbuettel D (2013).
\newblock \enquote{Fast and {{Elegant Numerical Linear Algebra Using}} the {{RcppEigen Package}}.}
\newblock \emph{Journal of Statistical Software}, \textbf{52}, 1--24.
\newblock ISSN 1548-7660.
\newblock \doi{10.18637/jss.v052.i05}.

\bibitem[{Baudin \emph{et~al.}(2016)Baudin, Dutfoy, Iooss, and Popelin}]{openTURNS_package}
Baudin M, Dutfoy A, Iooss B, Popelin AL (2016).
\newblock \emph{\pkg{OpenTURNS}: An Industrial Software for Uncertainty Quantification in Simulation}, pp. 1--38.
\newblock Springer International Publishing, Cham.
\newblock ISBN 978-3-319-11259-6.
\newblock \doi{10.1007/978-3-319-11259-6_64-1}.
\newblock \urlprefix\url{https://doi.org/10.1007/978-3-319-11259-6_64-1}.

\bibitem[{Bengtsson(2021)}]{future_package}
Bengtsson H (2021).
\newblock \enquote{A Unifying Framework for Parallel and Distributed Processing in \proglang{R} using Futures.}
\newblock \emph{The \proglang{R} Journal}, \textbf{13}(2), 208--227.
\newblock \doi{10.32614/RJ-2021-048}.
\newblock \urlprefix\url{https://doi.org/10.32614/RJ-2021-048}.

\bibitem[{Bidot \emph{et~al.}(2018)Bidot, Lamboni, and Monod}]{multisensi_package}
Bidot C, Lamboni M, Monod H (2018).
\newblock \emph{\pkg{multisensi}: Multivariate Sensitivity Analysis}.
\newblock \proglang{R} package version 2.1-1, \urlprefix\url{https://CRAN.R-project.org/package=multisensi}.

\bibitem[{Bonneel \emph{et~al.}(2011)Bonneel, {van de Panne}, Paris, and Heidrich}]{bonneelDisplacementInterpolationUsing2011a}
Bonneel N, {van de Panne} M, Paris S, Heidrich W (2011).
\newblock \enquote{Displacement Interpolation Using {{Lagrangian}} Mass Transport.}
\newblock In \emph{Proceedings of the 2011 {{SIGGRAPH Asia Conference}}}, {{SA}} '11, pp. 1--12. Association for Computing Machinery, New York, NY, USA.
\newblock ISBN 978-1-4503-0807-6.
\newblock \doi{10.1145/2024156.2024192}.

\bibitem[{Borgonovo(2007)}]{borgonovo2007new}
Borgonovo E (2007).
\newblock \enquote{A new uncertainty importance measure.}
\newblock \emph{Reliability Engineering \& System Safety}, \textbf{92}(6), 771--784.

\bibitem[{Borgonovo \emph{et~al.}(2025)Borgonovo, Figalli, Ghosal, Plischke, and Savar{\'e}}]{borgonovoConvexityMeasuresStatistical2025}
Borgonovo E, Figalli A, Ghosal P, Plischke E, Savar{\'e} G (2025).
\newblock \enquote{Convexity and Measures of Statistical Association.}
\newblock \emph{Journal of the Royal Statistical Society Series B: Statistical Methodology}, p. qkaf018.
\newblock ISSN 1369-7412.
\newblock \doi{10.1093/jrsssb/qkaf018}.

\bibitem[{Borgonovo \emph{et~al.}(2024)Borgonovo, Figalli, Plischke, and Savar{\'e}}]{borgonovoGlobalSensitivityAnalysis2024}
Borgonovo E, Figalli A, Plischke E, Savar{\'e} G (2024).
\newblock \enquote{Global {{Sensitivity Analysis}} via {{Optimal Transport}}.}
\newblock \emph{Management Science}.
\newblock \doi{10.1287/mnsc.2023.01796}.

\bibitem[{Borgonovo \emph{et~al.}(2016)Borgonovo, Hazen, and Plischke}]{borgonovoCommonRationaleGlobal2016}
Borgonovo E, Hazen GB, Plischke E (2016).
\newblock \enquote{A {{Common Rationale}} for {{Global Sensitivity Measures}} and {{Their Estimation}}.}
\newblock \emph{Risk Analysis}, \textbf{36}(10), 1871--1895.
\newblock ISSN 1539-6924.
\newblock \doi{10.1111/risa.12555}.

\bibitem[{Carlier \emph{et~al.}(2023)Carlier, Pegon, and Tamanini}]{carlier2023convergence}
Carlier G, Pegon P, Tamanini L (2023).
\newblock \enquote{Convergence rate of general entropic optimal transport costs.}
\newblock \emph{Calculus of Variations and Partial Differential Equations}, \textbf{62}(4), 116.

\bibitem[{Chatterjee(2021)}]{chatterjeeNewCoefficientCorrelation2021}
Chatterjee S (2021).
\newblock \enquote{A {{New Coefficient}} of {{Correlation}}.}
\newblock \emph{Journal of the American Statistical Association}, \textbf{116}(536), 2009--2022.
\newblock ISSN 0162-1459.
\newblock \doi{10.1080/01621459.2020.1758115}.

\bibitem[{Chatterjee and Holmes(2023)}]{xicor_package}
Chatterjee S, Holmes S (2023).
\newblock \emph{\pkg{XICOR}: Robust and generalized correlation coefficients}.
\newblock Https://github.com/spholmes/XICOR, \urlprefix\url{https://CRAN.R-project.org/package=XICOR}.

\bibitem[{Cuturi(2013)}]{cuturiSinkhornDistancesLightspeed2013}
Cuturi M (2013).
\newblock \enquote{Sinkhorn {{Distances}}: {{Lightspeed Computation}} of {{Optimal Transport}}.}
\newblock In \emph{Advances in {{Neural Information Processing Systems}}}, volume~26. Curran Associates, Inc.

\bibitem[{Cuturi \emph{et~al.}(2022)Cuturi, Meng-Papaxanthos, Tian, Bunne, Davis, and Teboul}]{OTT_package}
Cuturi M, Meng-Papaxanthos L, Tian Y, Bunne C, Davis G, Teboul O (2022).
\newblock \enquote{Optimal Transport Tools (\pkg{OTT}): A JAX Toolbox for all things Wasserstein.}
\newblock \emph{arXiv preprint arXiv:2201.12324}.

\bibitem[{Deb \emph{et~al.}(2020)Deb, Ghosal, and Sen}]{debMeasuringAssociationTopological2020}
Deb N, Ghosal P, Sen B (2020).
\newblock \enquote{Measuring {{Association}} on {{Topological Spaces Using Kernels}} and {{Geometric Graphs}}.}
\newblock \doi{10.48550/arXiv.2010.01768}.
\newblock \eprint{2010.01768}.

\bibitem[{Dunipace(2024)}]{approxOT_package}
Dunipace EA (2024).
\newblock \emph{\pkg{approxOT}: approximate optimal transport}.
\newblock \proglang{R} package version 1.1, \urlprefix\url{https://github.com/ericdunipace/approxOT}.

\bibitem[{Eddelbuettel and Francois(2011)}]{eddelbuettelRcppSeamlessIntegration2011}
Eddelbuettel D, Francois R (2011).
\newblock \enquote{\pkg{Rcpp}: {{Seamless \proglang{R}}} and \proglang{C++} {{Integration}}.}
\newblock \emph{Journal of Statistical Software}, \textbf{40}, 1--18.
\newblock ISSN 1548-7660.
\newblock \doi{10.18637/jss.v040.i08}.

\bibitem[{Figalli and Glaudo(2021)}]{figalliInvitationOptimalTransport2021}
Figalli A, Glaudo F (2021).
\newblock \enquote{An {{Invitation}} to {{Optimal Transport}}, {{Wasserstein Distances}}, and {{Gradient Flows}}.}
\newblock https://ems.press/books/etb/190.
\newblock \doi{10.4171/etb/22}.

\bibitem[{Flamary \emph{et~al.}(2021)Flamary, Courty, Gramfort, Alaya, Boisbunon, Chambon, Chapel, Corenflos, Fatras, Fournier, Gautheron, Gayraud, Janati, Rakotomamonjy, Redko, Rolet, Schutz, Seguy, Sutherland, Tavenard, Tong, and Vayer}]{flamary2021pot}
Flamary R, Courty N, Gramfort A, Alaya MZ, Boisbunon A, Chambon S, Chapel L, Corenflos A, Fatras K, Fournier N, Gautheron L, Gayraud NT, Janati H, Rakotomamonjy A, Redko I, Rolet A, Schutz A, Seguy V, Sutherland DJ, Tavenard R, Tong A, Vayer T (2021).
\newblock \enquote{\pkg{POT}: Python Optimal Transport.}
\newblock \emph{Journal of Machine Learning Research}, \textbf{22}(78), 1--8.
\newblock \urlprefix\url{http://jmlr.org/papers/v22/20-451.html}.

\bibitem[{Gamboa \emph{et~al.}(2014)Gamboa, Janon, Klein, and Lagnoux}]{gamboaSensitivityAnalysisMultidimensional2014}
Gamboa F, Janon A, Klein T, Lagnoux A (2014).
\newblock \enquote{Sensitivity Analysis for Multidimensional and Functional Outputs.}
\newblock \emph{Electronic Journal of Statistics}, \textbf{8}(1), 575--603.
\newblock ISSN 1935-7524, 1935-7524.
\newblock \doi{10.1214/14-EJS895}.

\bibitem[{Gelbrich(1990)}]{gelbrichFormulaL2Wasserstein1990}
Gelbrich M (1990).
\newblock \enquote{On a {{Formula}} for the {{L2 Wasserstein Metric}} between {{Measures}} on {{Euclidean}} and {{Hilbert Spaces}}.}
\newblock \emph{Mathematische Nachrichten}, \textbf{147}(1), 185--203.
\newblock ISSN 1522-2616.
\newblock \doi{10.1002/mana.19901470121}.

\bibitem[{Gottschlich and Schuhmacher(2014)}]{gottschlichShortlistMethodFast2014}
Gottschlich C, Schuhmacher D (2014).
\newblock \enquote{The {{Shortlist Method}} for {{Fast Computation}} of the {{Earth Mover}}'s {{Distance}} and {{Finding Optimal Solutions}} to {{Transportation Problems}}.}
\newblock \emph{PLOS ONE}, \textbf{9}(10), e110214.
\newblock ISSN 1932-6203.
\newblock \doi{10.1371/journal.pone.0110214}.

\bibitem[{Herman and Usher(2017)}]{SALib_package_2}
Herman J, Usher W (2017).
\newblock \enquote{\pkg{SALib}: An open-source \proglang{Python} library for Sensitivity Analysis.}
\newblock \emph{The Journal of Open Source Software}, \textbf{2}(9).
\newblock \doi{10.21105/joss.00097}.
\newblock \urlprefix\url{https://doi.org/10.21105/joss.00097}.

\bibitem[{Iooss \emph{et~al.}(2022)Iooss, Veiga, Janon, Pujol, with contributions~from Baptiste~Broto, Boumhaout, Delage, Amri, Fruth, Gilquin, Guillaume, Herin, Idrissi, {Le Gratiet}, Lemaitre, Marrel, Meynaoui, Nelson, Monari, Oomen, Rakovec, Ramos, Roustant, Sarazin, Song, Staum, Sueur, Touati, Verges, and Weber}]{sensitivity_package}
Iooss B, Veiga SD, Janon A, Pujol G, with contributions~from Baptiste~Broto, Boumhaout K, Delage T, Amri RE, Fruth J, Gilquin L, Guillaume J, Herin M, Idrissi MI, {Le Gratiet} L, Lemaitre P, Marrel A, Meynaoui A, Nelson BL, Monari F, Oomen R, Rakovec O, Ramos B, Roustant O, Sarazin G, Song E, Staum J, Sueur R, Touati T, Verges V, Weber F (2022).
\newblock \emph{\pkg{sensitivity}: Global Sensitivity Analysis of Model Outputs}.
\newblock \proglang{R} package version 1.28.0, \urlprefix\url{https://CRAN.R-project.org/package=sensitivity}.

\bibitem[{Iwanaga \emph{et~al.}(2022)Iwanaga, Usher, and Herman}]{SALib_package_1}
Iwanaga T, Usher W, Herman J (2022).
\newblock \enquote{Toward \pkg{SALib} 2.0: {Advancing} the accessibility and interpretability of global sensitivity analyses.}
\newblock \emph{Socio-Environmental Systems Modelling}, \textbf{4}, 18155.
\newblock \doi{10.18174/sesmo.18155}.
\newblock \urlprefix\url{https://sesmo.org/article/view/18155}.

\bibitem[{Janati \emph{et~al.}(2020)Janati, Muzellec, Peyr{\'e}, and Cuturi}]{janatiEntropicOptimalTransport2020}
Janati H, Muzellec B, Peyr{\'e} G, Cuturi M (2020).
\newblock \enquote{Entropic {{Optimal Transport}} between {{Unbalanced Gaussian Measures}} Has a {{Closed Form}}.}
\newblock In \emph{Advances in {{Neural Information Processing Systems}}}, volume~33, pp. 10468--10479. Curran Associates, Inc.

\bibitem[{Lamboni \emph{et~al.}(2011)Lamboni, Monod, and Makowski}]{lamboniMultivariateSensitivityAnalysis2011}
Lamboni M, Monod H, Makowski D (2011).
\newblock \enquote{Multivariate Sensitivity Analysis to Measure Global Contribution of Input Factors in Dynamic Models.}
\newblock \emph{Reliability Engineering \& System Safety}, \textbf{96}(4), 450--459.
\newblock ISSN 0951-8320.
\newblock \doi{10.1016/j.ress.2010.12.002}.

\bibitem[{Ludwig \emph{et~al.}(2022)Ludwig, Jones, and Holling}]{ludwigQualitativeAnalysisInsect2022}
Ludwig D, Jones DD, Holling CS (2022).
\newblock \enquote{Qualitative Analysis of Insect Outbreak Systems: The Spruce Budworm and Forest.}
\newblock In \emph{Qualitative Analysis of Insect Outbreak Systems: The Spruce Budworm and Forest}, pp. 547--564. University of Chicago Press.
\newblock ISBN 978-0-226-12553-4.
\newblock \doi{10.7208/chicago/9780226125534-035}.

\bibitem[{Luenberger and Ye(2021)}]{luenbergerLinearNonlinearProgramming2021}
Luenberger DG, Ye Y (2021).
\newblock \emph{Linear and {{Nonlinear Programming}}}, volume 228 of \emph{International {{Series}} in {{Operations Research}} \& {{Management Science}}}.
\newblock Springer International Publishing, Cham.
\newblock ISBN 978-3-030-85449-2 978-3-030-85450-8.
\newblock \doi{10.1007/978-3-030-85450-8}.

\bibitem[{Marelli and Sudret(2014)}]{marelliUQLabFrameworkUncertainty2014}
Marelli S, Sudret B (2014).
\newblock \enquote{{{UQLab}}: {{A Framework}} for {{Uncertainty Quantification}} in {{\proglang{MATLAB}}}.}
\newblock pp. 2554--2563.
\newblock \doi{10.1061/9780784413609.257}.

\bibitem[{M{\'o}ri and Sz{\'e}kely(2019)}]{moriFourSimpleAxioms2019a}
M{\'o}ri TF, Sz{\'e}kely GJ (2019).
\newblock \enquote{Four Simple Axioms of Dependence Measures.}
\newblock \emph{Metrika}, \textbf{82}(1), 1--16.
\newblock ISSN 1435-926X.
\newblock \doi{10.1007/s00184-018-0670-3}.

\bibitem[{M{\'o}ri and Sz{\'e}kely(2020)}]{moriEarthMoverCorrelation2020}
M{\'o}ri TF, Sz{\'e}kely GJ (2020).
\newblock \enquote{The {{Earth Mover}}'s {{Correlation}}.}
\newblock \doi{10.48550/arXiv.2009.04313}.
\newblock \eprint{2009.04313}.

\bibitem[{Nies \emph{et~al.}(2023)Nies, Staudt, and Munk}]{niesTransportDependencyOptimal2023}
Nies TG, Staudt T, Munk A (2023).
\newblock \enquote{Transport {{Dependency}}: {{Optimal Transport Based Dependency Measures}}.}
\newblock \doi{10.48550/arXiv.2105.02073}.
\newblock \eprint{2105.02073}.

\bibitem[{Noacco \emph{et~al.}(2019)Noacco, Sarrazin, Pianosi, and Wagener}]{noacco2019matlab}
Noacco V, Sarrazin F, Pianosi F, Wagener T (2019).
\newblock \enquote{\proglang{MATLAB}/\proglang{R} workflows to assess critical choices in Global Sensitivity Analysis using the SAFE toolbox.}
\newblock \emph{MethodsX}, \textbf{6}, 2258--2280.

\bibitem[{Nordhaus(2017)}]{nordhausRevisitingSocialCost2017}
Nordhaus WD (2017).
\newblock \enquote{Revisiting the Social Cost of Carbon.}
\newblock \emph{Proceedings of the National Academy of Sciences}, \textbf{114}(7), 1518--1523.
\newblock \doi{10.1073/pnas.1609244114}.

\bibitem[{Pearson(1905)}]{pearsonGeneralTheorySkew1905}
Pearson K (1905).
\newblock \enquote{On the {{General Theory}} of {{Skew Correlation}} and {{Non-Linear Regression}}.}

\bibitem[{Pedersen(2024)}]{patchwork_package}
Pedersen TL (2024).
\newblock \emph{\pkg{patchwork}: The Composer of Plots}.
\newblock \proglang{R} package version 1.2.0, \urlprefix\url{https://CRAN.R-project.org/package=patchwork}.

\bibitem[{Peyr{\'e} and Cuturi(2020)}]{peyreComputationalOptimalTransport2020}
Peyr{\'e} G, Cuturi M (2020).
\newblock \enquote{Computational {{Optimal Transport}}.}
\newblock \doi{10.48550/arXiv.1803.00567}.
\newblock \eprint{1803.00567}.

\bibitem[{Pianosi \emph{et~al.}(2015)Pianosi, Sarrazin, and Wagener}]{pianosiMatlabToolboxGlobal2015}
Pianosi F, Sarrazin F, Wagener T (2015).
\newblock \enquote{A {{\proglang{MATLAB}}} Toolbox for {{Global Sensitivity Analysis}}.}
\newblock \emph{Environmental Modelling \& Software}, \textbf{70}, 80--85.
\newblock ISSN 1364-8152.
\newblock \doi{10.1016/j.envsoft.2015.04.009}.

\bibitem[{Pianosi and Wagener(2015)}]{pianosi2015simple}
Pianosi F, Wagener T (2015).
\newblock \enquote{A simple and efficient method for global sensitivity analysis based on cumulative distribution functions.}
\newblock \emph{Environmental Modelling \& Software}, \textbf{67}, 1--11.

\bibitem[{Puy \emph{et~al.}(2022)Puy, Piano, Saltelli, and Levin}]{puySensobolPackageCompute2022}
Puy A, Piano SL, Saltelli A, Levin SA (2022).
\newblock \enquote{\pkg{sensobol}: {{An \proglang{R} Package}} to {{Compute Variance-Based Sensitivity Indices}}.}
\newblock \emph{Journal of Statistical Software}, \textbf{102}, 1--37.
\newblock ISSN 1548-7660.
\newblock \doi{10.18637/jss.v102.i05}.

\bibitem[{Razavi \emph{et~al.}(2021)Razavi, Jakeman, Saltelli, Prieur, Iooss, Borgonovo, Plischke, Lo~Piano, Iwanaga, Becker, Tarantola, Guillaume, Jakeman, Gupta, Melillo, Rabitti, Chabridon, Duan, Sun, Smith, Sheikholeslami, Hosseini, Asadzadeh, Puy, Kucherenko, and Maier}]{razaviFutureSensitivityAnalysis2021}
Razavi S, Jakeman A, Saltelli A, Prieur C, Iooss B, Borgonovo E, Plischke E, Lo~Piano S, Iwanaga T, Becker W, Tarantola S, Guillaume JHA, Jakeman J, Gupta H, Melillo N, Rabitti G, Chabridon V, Duan Q, Sun X, Smith S, Sheikholeslami R, Hosseini N, Asadzadeh M, Puy A, Kucherenko S, Maier HR (2021).
\newblock \enquote{The {{Future}} of {{Sensitivity Analysis}}: {{An}} Essential Discipline for Systems Modeling and Policy Support.}
\newblock \emph{Environmental Modelling \& Software}, \textbf{137}, 104954.
\newblock ISSN 1364-8152.
\newblock \doi{10.1016/j.envsoft.2020.104954}.

\bibitem[{R{\'e}nyi(1959)}]{renyiMeasuresDependence1959a}
R{\'e}nyi A (1959).
\newblock \enquote{On Measures of Dependence.}
\newblock \emph{Acta Mathematica Academiae Scientiarum Hungarica}, \textbf{10}(3), 441--451.
\newblock ISSN 1588-2632.
\newblock \doi{10.1007/BF02024507}.

\bibitem[{Reshef \emph{et~al.}(2011)Reshef, Reshef, Finucane, Grossman, McVean, Turnbaugh, Lander, Mitzenmacher, and Sabeti}]{reshef2011detecting}
Reshef DN, Reshef YA, Finucane HK, Grossman SR, McVean G, Turnbaugh PJ, Lander ES, Mitzenmacher M, Sabeti PC (2011).
\newblock \enquote{Detecting novel associations in large data sets.}
\newblock \emph{science}, \textbf{334}(6062), 1518--1524.

\bibitem[{Saltelli \emph{et~al.}(2008)Saltelli, Ratto, Andres, Campolongo, Cariboni, Gatelli, Saisana, and Tarantola}]{saltelliGlobalSensitivityAnalysis2008}
Saltelli A, Ratto M, Andres T, Campolongo F, Cariboni J, Gatelli D, Saisana M, Tarantola S (2008).
\newblock \emph{Global {{Sensitivity Analysis}}: {{The Primer}}}.
\newblock John Wiley \& Sons.
\newblock ISBN 978-0-470-72517-7.

\bibitem[{Santambrogio(2015)}]{santambrogioOptimalTransportApplied}
Santambrogio F (2015).
\newblock \emph{Optimal {{Transport}} for {{Applied Mathematicians}}}.
\newblock Birkh\"auser, Cham.

\bibitem[{Schmitzer(2019)}]{schmitzerStabilizedSparseScaling2019}
Schmitzer B (2019).
\newblock \enquote{Stabilized {{Sparse Scaling Algorithms}} for {{Entropy Regularized Transport Problems}}.}
\newblock \emph{SIAM Journal on Scientific Computing}, \textbf{41}(3), A1443--A1481.
\newblock ISSN 1064-8275.
\newblock \doi{10.1137/16M1106018}.

\bibitem[{Schuhmacher \emph{et~al.}(2024)Schuhmacher, Bähre, Bonneel, Gottschlich, Hartmann, Heinemann, Schmitzer, and Schrieber}]{transport_package}
Schuhmacher D, Bähre B, Bonneel N, Gottschlich C, Hartmann V, Heinemann F, Schmitzer B, Schrieber J (2024).
\newblock \enquote{\pkg{transport}: Computation of Optimal Transport Plans and Wasserstein Distances.}
\newblock \proglang{R} package version 0.15-2, \urlprefix\url{https://cran.r-project.org/package=transport}.

\bibitem[{Sinkhorn(1967)}]{sinkhornDiagonalEquivalenceMatrices1967}
Sinkhorn R (1967).
\newblock \enquote{Diagonal {{Equivalence}} to {{Matrices}} with {{Prescribed Row}} and {{Column Sums}}.}
\newblock \emph{The American Mathematical Monthly}, \textbf{74}(4), 402--405.
\newblock ISSN 0002-9890.
\newblock \doi{10.2307/2314570}.
\newblock \eprint{2314570}.

\bibitem[{Sobol(2001)}]{sobol2001global}
Sobol IM (2001).
\newblock \enquote{Global sensitivity indices for nonlinear mathematical models and their Monte Carlo estimates.}
\newblock \emph{Mathematics and computers in simulation}, \textbf{55}(1-3), 271--280.

\bibitem[{Soetaert \emph{et~al.}(2010)Soetaert, Petzoldt, and Setzer}]{desolve_package}
Soetaert K, Petzoldt T, Setzer RW (2010).
\newblock \enquote{Solving Differential Equations in {\proglang{R}}: Package \pkg{de{S}olve}.}
\newblock \emph{Journal of Statistical Software}, \textbf{33}(9), 1--25.
\newblock \doi{10.18637/jss.v033.i09}.

\bibitem[{Strong and Oakley(2013)}]{strong2013efficient}
Strong M, Oakley JE (2013).
\newblock \enquote{An efficient method for computing partial expected value of perfect information for correlated inputs.}
\newblock \emph{Medical Decision-Making}, \textbf{33}(6), 755--766.

\bibitem[{Vaughan and Dancho(2022)}]{furrr_package}
Vaughan D, Dancho M (2022).
\newblock \emph{\pkg{furrr}: Apply Mapping Functions in Parallel using Futures}.
\newblock \proglang{R} package version 0.3.1, \urlprefix\url{https://CRAN.R-project.org/package=furrr}.

\bibitem[{Villani(2009)}]{villaniOptimalTransport2009}
Villani C (2009).
\newblock \emph{Optimal {{Transport}}}, volume 338 of \emph{Grundlehren Der Mathematischen {{Wissenschaften}}}.
\newblock Springer, Berlin, Heidelberg.
\newblock ISBN 978-3-540-71049-3 978-3-540-71050-9.
\newblock \doi{10.1007/978-3-540-71050-9}.

\bibitem[{Wickham(2016)}]{ggplot2_package}
Wickham H (2016).
\newblock \emph{\pkg{ggplot2}: Elegant Graphics for Data Analysis}.
\newblock Springer-Verlag New York.
\newblock ISBN 978-3-319-24277-4.
\newblock \urlprefix\url{https://ggplot2.tidyverse.org}.

\bibitem[{Wiesel(2022)}]{wieselMeasuringAssociationWasserstein2022}
Wiesel JCW (2022).
\newblock \enquote{Measuring Association with {{Wasserstein}} Distances.}
\newblock \emph{Bernoulli}, \textbf{28}(4), 2816--2832.
\newblock ISSN 1350-7265.
\newblock \doi{10.3150/21-BEJ1438}.

\bibitem[{You(2023)}]{T4transport_package}
You K (2023).
\newblock \emph{\pkg{T4transport}: Tools for Computational Optimal Transport}.
\newblock \proglang{R} package version 0.1.2, \urlprefix\url{https://CRAN.R-project.org/package=T4transport}.

\end{thebibliography}


\newpage

\begin{appendix}

\end{appendix}


\end{document}